\documentclass[nofootinbib,preprintnumbers,amsmath,amssymb, 11pt]{article}
\usepackage[english]{babel}
\usepackage[a4paper, inner=2cm, outer=2cm, top=3cm, bottom=3cm]{geometry}
\usepackage[dvipsnames]{xcolor}
\usepackage{mathtools}
\usepackage{pstricks}
\usepackage{pst-node}
\usepackage{pst-plot}
\usepackage{caption}
\usepackage{graphicx}
\usepackage{amsmath}
\usepackage{amssymb}
\usepackage{mathcomp}
\usepackage{cancel}
\usepackage{textcomp}
\usepackage{enumitem}
\usepackage{physics}
\usepackage{romannum}
\usepackage[doublespacing]{setspace}
\usepackage{subcaption}
\usepackage[linktoc=all]{hyperref}
\usepackage{booktabs}
\usepackage{cite}
\usepackage[utf8]{inputenc}
\usepackage{authblk}
\usepackage{bm}
\usepackage{array}
\usepackage{graphicx}
\usepackage{caption}
\usepackage{lipsum}
\usepackage{empheq}
\usepackage[many]{tcolorbox}
\usepackage[normalem]{ulem}
\usepackage{float}
\usepackage{gensymb}

\hypersetup{
  colorlinks   = true,
  urlcolor     = Blue, 
  linkcolor    = Magenta,
  citecolor    = Magenta
}

\tcbset{highlight math style={enhanced,
colframe=black,colback=white,arc=2mm,boxrule=0.9pt}}

\newcommand{\Ddv}[2]{\frac{\mathrm{D} #1}{\mathrm{D} #2}}
\newcommand{\LDdv}[2]{\mathrm{D} #1 / \mathrm{D} #2}
\newcommand{\BM}{{\tiny\mathtt{B}}}

\title{On the Emergence of the Hysteresis Effect in Gravitational Systems}
\author[1]{Raihaneh Moti\thanks{\href{mailto:r.moti@ipm.ir}{\ r.moti@ipm.ir}}}
\author[2]{Ali Shojai\thanks{\href{mailto:ashojai@ut.ac.ir}{\ ashojai@ut.ac.ir}  (Corresponding Author)}}
\affil[1]{{\small\textit{School of Astronomy, Institute for Research in Fundamental Sciences (IPM), \hspace{4cm} Tehran, Iran P.O. Box 19395-5531}} }
\affil[2]{{\small\textit{Department of Physics, University of Tehran, North Karegar St., Tehran, Iran}}}

\date{}

\begin{document}
\maketitle
\pagenumbering{arabic}

\begin{abstract}
Previously, it was shown that the interplay between the gravitational memory effect and the thermodynamics of an ensemble of spinning objects gives rise to a hysteresis effect, which can be termed {\it Gravo-Thermo memory}. 
This effect can be studied either through a statistical approach or via kinetic theory. 
Here, we present a general kinetic theory model for the general case of a fluid composed of spinning extended objects in the presence of gravity. 
\end{abstract}

\section{Introduction}
\label{Sec:Intro.}

Hysteresis effect emerges when the initial state of a system cannot be recovered simply by reversing the direction of the evolution, and the system retains some memory of its past history. As a result, it introduces an effective arrow of evolution for the system.
Hysteresis is usually associated with either non-linear dynamics of open systems (like the magnetic hysteresis of materials), where the superposition principle generally fails, or statistical representations of macroscopic systems (like any irreversible thermodynamic process). 

Although higher order quantum effects can introduce non-linear terms (like in quantum electrodynamics), for the case of gravity and the gravitational systems, the problem is slightly different.  Einstein gravity is intrinsically non-linear at the classical level since the gravitational field itself acts as a source of gravity. In addition, general relativity is a theory of spacetime and thus its non-linear character can lead to the emergence of some interconnection of solutions at some time to the past history of the solution. These two inherent properties underlies a family of phenomena collectively known as gravitational memory effect \cite{Zeldovich:1974gvh, Braginsky:1987kwo, Christodoulou:1991cr, Thorne:1992sdb, Mitman:2024uss}.

Gravitational memory manifests itself as a permanent change in the state of a physical system and the gravity itself after its interaction with a gravitational field. Depending on the nature of the gravitational source and the properties of the system, memory may manifest through a variety of observables, including displacements between constituent particles\cite{Favata:2010zu, 
Mitman:2020pbt, Flanagan:2018yzh, Wang:2023eqj, Zhang:2024uyp, Zhang:2024tey}, persistent changes in their relative velocities\cite{Grishchuk:1989qa, Divakarla:2021xrd, Bieri:2024ios, BenAchour:2024ucn}, and modifications of their relative orientation and spin degrees of freedom\cite{Nichols:2017rqr, Pasterski:2015tva, Harte:2024mwj}.

This intrinsic memorization ability of gravity, can be mixed up with statistical nature of an ensemble of such systems to create a hysteresis effect originating both from non-linearity and thermodynamic arrow of evolution. One may ask whether memorized observables can be employed as statistical quantities, registering the footprint of gravitational interactions. Such \textit{Gravo-Thermo memory} effects are explored in \cite{Moti:2024a,Moti:2024b,Moti:2025}. The central question is whether observable thermodynamic quantities encode information about the system's dynamical history.

In general such an ensemble of gravitational systems, may consists of spinless/spinning point-like systems or spinless/spinning extended objects.
In this paper, we revisit the idea in detail and propose a general framework for investigating Gravo-Thermo memory phenomena. We see that it is possible to explore the effect either by using statistical mechanics or the kinetic theory. This can be done both for point-like and extended objects. 

The purpose of this paper is to put our previous works \cite{Moti:2024a,Moti:2024b,Moti:2025} on such an effect for point particles and extended objects, in a coherent framework and then provide a kinetic theoretical approach to the Gravo-Thermo memory of a flow of spinning extended objects in section \ref{SecIII:Kinetic}. To do so, we begin by studying gyroscopic memory for an ensemble of spinning point particles and subsequently extend the analysis to an ensembles of extended bodies whose dynamics are governed by the Mathisson--Papapetrou--Dixon equations rather than geodesic motion. 
We will discuss the main framework and present an overall perspective of the effect.

\section{An Ensemble of Spinning Point Particles}
\label{Sec:II}

As the first step in describing Gravo-Thermo memory effect, let's start with spinning point particles. To do so we need to note that the spin direction and its precession has memory during the gravitational interaction. 

Since the spin four-vector $S^{\mu}$ of a small freely falling gyroscope is parallel transported as
\begin{equation}
	v^{\nu} \nabla_{\nu} S^{\mu} = 0
\end{equation}
with $v^{\nu}$ the proper velocity along the geodesic, for a gyroscope with purely spatial orientation $S^{\hat{i}} = S^{\mu} e^{\hat{i}}_{\ \mu}$ and $S^{\hat{0}}=0$ in its local co-moving frame $e_{\hat{a}}^{\ \mu} = (v^{\mu},e_{\hat{i}}^{\ \mu}) $, we have 
\begin{equation}
\dv{S^{\hat{i}}}{\lambda} = \Omega^{\hat{i}}_{~ \hat{j}} (\lambda) S^{\hat{j}} (\lambda)
\label{Par-Tra}
\end{equation}
where $\lambda$ is the observer's proper time and the spacetime properties are encoded in the angular velocity  $ \Omega^{\hat{i}\hat{j}} = - v^{\alpha} \omega_{\alpha}^{~\hat{i}\hat{j}} $ as the projection of the spin connections of the curved spacetime $\omega_{\mu}^{~ \hat{\mu}\hat{\nu}} = e^{\hat{\mu}}_{~ \alpha}\nabla_{\mu}e^{~ \hat{\nu}\alpha}$ along $v^{\alpha}$ \cite{Book:MTW}.
Hence, the gyroscope precession rate is determined by the spin connections evaluated along the observer's worldline.

The Bondi coordinate system properly describes the gravitational radiation at the far regions from the source \cite{Madler:2016}. The retarded Bondi coordinates $x^{\alpha} = (u,r, \chi^{1},\chi^{2})$ are based on a family of outgoing null hypersurfaces with constant $u\equiv t-r$. While the coordinate $r$ varies  along the null rays, the angular coordinates $\chi^{A}=(\theta,\phi)$ remain constant along them. Then metric near the future null infinity, is
\begin{equation}
\dd{s}^2 = -Ue^{2\tilde{\beta}} \dd{u^2} - 2e^{2\tilde{\beta}} \dd{u} \dd{r} +r^2\gamma_{AB}\left(\dd{\chi^A}-\mathcal{U}^A\dd{u}\right)\left(\dd{\chi^B}-\mathcal{U}^B\dd{u}\right)
\label{Bondi}
\end{equation}
where $A,B=1,2$ and $U, \tilde{\beta}, \mathcal{U}^{A}$ and $\gamma_{AB}$ are functions of coordinates $(u,r,\chi^A)$. On the  null hypersurface at infinity, source oriented local frame is defined as the one with its timelike vector along the retarded time $u$.

Applying the parallel transport equation \eqref{Par-Tra} in the Bondi coordinate system \eqref{Bondi}, it was shown in \cite{Seraj:2021rxd, Seraj:2022qyt} that a gyroscope with an arbitrary initial orientation $S_{\hat{i}}(u_0)$ undergoes a net precession due to the passage of gravitational waves. The resulting change in its orientation is
\begin{equation}
\Delta S^{\hat{r}} = \mathcal{O}(r^{-3}) \quad \quad, \quad \quad \Delta S^{\hat{A}} = \epsilon^{\hat{A}\hat{B}} S_{\hat{B}}(u_0) \Delta\Phi + \mathcal{O}(r^{-3})
\label{Delta-S}
\end{equation}
where the accumulated precession angle is given by
\begin{equation}
\Delta\Phi =-\dfrac{1}{r^2} \int \dd{u} \left(\dfrac{1}{4} \nabla_A \nabla_B \tilde{C}^{AB} - \dfrac{1}{8} N_{AB} \tilde{C}^{AB} \right) \ .
\label{Precession}
\end{equation}
Here, $\tilde{C}_{AB}\equiv \epsilon_{AC} C_{B}^{\ C}$ is the dual of the Bondi shear $C_{AB}$, which measures the shear of a congruence of outgoing null rays beside encoding the gravitational waveform reaching null infinity at retarded time $u$ and angular coordinates $\chi^A$. The \textit{news tensor} $N_{AB}\equiv \partial_u C_{AB}$ determines time dependence of the Bondi shear and is a measure of the energy flux of the gravitational radiation.

Since $\Delta\Phi$ quantifies the permanent spin precession of point particles resulting from the gravitational interaction, it would be a good parameter for investigation of thermodynamic character of an ensemble of such spinning point particles. One can do this either by statistical mechanics approach or via kinetic theory as we shall see in the following subsections.

\subsection{Statistical Description}
\label{SecII:Thermo}

Now consider an ensemble of spinning particles confined within a small freely falling box, and suppose that the passage of a gravitational wave perturbs the motion of each particle, leading to a change in their spinning orientation with respect to the inertial frame defined by distant fixed stars. If the ensemble is initially prepared in a state of partial spin alignment, neither perfectly ordered nor completely random, then the imprint of the gravitational interaction is encoded in the final statistical state of the system. This effect is expected to manifest itself in the partition function describing the final equilibrium state of the ensemble, such that \cite{Moti:2024a}
\begin{equation}
\Delta\mathcal{Z} \equiv \mathcal{Z}_f - \mathcal{Z}_i \propto \int e^{-\beta H_{\text{int}} (\Delta S)} \dd{\Omega}
\end{equation}
where
$ \mathcal{Z}_{f/i} = \int e^{-\beta H_{f/i} } \dd{\Omega}$, 
and $ H_{\text{int}} (\Delta S) = H_f - H_i $.

To construct a more practical model, consider an ensemble of paramagnetic particles located in a uniform external magnetic field $\vec{B}$. Each particle with initial spin $\vec{S}_0$ carries a magnetic moment $\mu$ and interacts with the field through the Hamiltonian $ H_0 = -\mu \vec{B}\vdot\vec{S}_0$. The passage of a gravitational pulse induces a spin precession, resulting in a change  $\Delta \vec{S}$  of the spin orientation. Without loss of generality, we assume that the magnetic field lies entirely within the angular $\chi^A$ surface and is therefore orthogonal to the radial direction $\vec{r}$. For justification of this assumption see \ref{App. A}.

The angle between the magnetic field $\vec{B}$ and the initial spin orientation is denoted by $\vartheta$.
According to the precession memory effect \eqref{Precession}, the projection of $\Delta\vec{S}$ onto the angular-surface is orthogonal to the corresponding projection of the initial spin vector $\vec{S}_0$. Consequently, the induced spin displacement $\Delta\vec{S}$ forms an angle $ \vartheta' \equiv \frac{\pi}{2}-\vartheta $ with the magnetic field. For a uniform magnetic field with fixed magnitude, $B \equiv \abs*{\vec{B}}=\mathrm{const.}$, these considerations leads to \cite{Moti:2024a}
\begin{equation}
\mathcal{Z} =  2\pi \int e^{\beta\mu B S_0 \cos{\vartheta}}\ \times  e^{\beta\mu B S_0 \Delta\Phi \sin{\vartheta}}  \dd{(\cos{\vartheta})}
\label{NonP-PF}
\end{equation}
where $S_0 \equiv \abs*{\vec{S_0}}$.
Since $\Delta\Phi \propto -1/r^2$,  far from the source we can use the approximated relation
\begin{equation}
\mathcal{Z}  \simeq 2\pi \int e^{\beta\chi \cos{\vartheta}}  \bigl(1+  \beta\chi\Delta\Phi \sin{\vartheta}  \bigr) \dd{(\cos{\vartheta})} \equiv \mathcal{Z}^{(0)}+\Delta\Phi \tilde{\mathcal{Z}}
\label{PF}
\end{equation}
in which $ \chi \equiv \mu B S_0$ and $\mathcal{Z}^{(0)}$ is the initial memory-less term. 
Since $\Delta\Phi$ is small at large distances from the gravitational source, the memory contribution to all thermodynamic quantities can be treated perturbatively. To leading order, the correction is proportional to $\Delta\Phi$ and originates from the memory-dependent term $\tilde{\mathcal{Z}}$.

As it is shown in \cite{Moti:2024a}, using this one can estimate and predict physical effects of the passage of a gravitational wave pulse on the thermodynamic quantities of such an ensemble of spinning point particles. 
For example, one can estimate the magnitude of the effect on the specific heat for Tungsten as a typical paramagnetic material (magnetic susceptibility of $6.8\times10^{-5}$), placed in an external magnetic field of strength $100-1000$ Gauss. As the gravitational wave source, they \cite{Moti:2024a} consider a binary neutron star (BNS) system with component masses in the observationally allowed range $1.1 M_{\odot} < m < 1.7 M_{\odot}$, rotation periods between $0.1 \text{ day}$ to $45 \text{ days}$ and distance of order $10\text{ Mpc}$. Then, a fractional change in the specific heat of approximately $\dfrac{\Delta C_v}{C_v} \sim 10^{-7}-10^{-6}$ per gravitational wave pulse is obtained.

Consequently, if the temperature of one kilogram of Tungsten is increased by $10$ degrees, the gravitational memory effect changes the required energy by approximately $10 \mu J$ to $10 mJ$ per pulse. Such an effect is, in principle, measurable, provided that the experimental apparatus is sufficiently isolated from environmental disturbances and can be aligned with the required precision.

\subsection{Kinetic Theoretical Description}
\label{SecII:Kinetic}

The same problem may also be formulated using the kinetic theory in curved spacetime, which has the advantage of describing the system in terms of average dynamical quantities through the Boltzmann equation. This would provide an appropriate framework for investigating the corresponding effects in a flowing medium rather than in a finite ensemble of stationary particles. 

The macroscopic observables of the system are obtained from the distribution function $f \equiv f(x^{\mu}, p^{\mu})$, whose dynamics is  governed by the Boltzmann equation. In curved spacetime, gravitational effects enter through the connection coefficients, leading to the modified Boltzmann equation \cite{Cercignani:2002}
\begin{equation}
p^{\mu} \pdv{f}{x^{\mu}} - \Gamma^{i}_{\mu\nu} p^{\mu} p^{\nu} \pdv{f}{p^{i}} = Q
\end{equation}
where $Q \equiv Q(f,F,\Omega,p)$ contains collision effects in which $\Omega$ is the solid angle element characterizing binary collisions between the particles of the fluid and $F$ is the invariant particle flux. (See Appendix of reference \cite{Moti:2024b})

Obtaining the exact solution of this equation for the distribution function $f$ is always a challenging task and it is better to explore the macroscopic behavior of the system by obtaining the governing dynamical equations of main kinetic moments. Motivated by this, one can focus on three essential moments: particle number current $\mathcal{N}^{\mu}_{\BM}$, energy--momentum tensor $\mathcal{T}^{\mu\nu}_{\BM}$ and entropy current $\zeta^{\mu}_{\BM}$ which are defined as\footnote{The subscript $\BM$ is introduced to distinguish these essential moments of the Boltzmann distribution function from the moments describing the matter distribution of extended objects, used in the next section.}
\begin{align}
      & \mathcal{N}^{\mu}_{\BM} = c \displaystyle\int f p^{\mu}  \sqrt{g} \dfrac{\dd[3]p}{p_0} \label{PN-Mom}\\
      & \mathcal{T}^{\mu\nu}_{\BM}  = c \displaystyle\int f p^{\mu} p^{\nu}  \sqrt{g} \dfrac{\dd[3]p}{p_0} \label{EMT-Mom}\\
      & \zeta^{\mu}_{\BM} = -ck  \displaystyle\int  f\ \ln{\left (\dfrac{f h^3}{e g_s}\right )}\ p^{\mu}  \sqrt{g} \dfrac{\dd[3]{p}}{p_0}  \label{Ent-Mom} 
\end{align}
where $h$, $e$, and $g_s$ are Planck constant, internal energy per particle and spin degeneracy factor, respectively.

From the Boltzmann equation, it  is shown \cite{Cercignani:2002} that these three quantities satisfy
\begin{align}
      &  \mathcal{N}^{\mu}_{\BM \ ;\mu} = 0  \\
      &  \mathcal{T}^{\mu\nu}_{\BM \ \ ;\mu}  = 0\\
      &  \zeta^{\mu}_{\BM \ ;\mu} \geq 0
\end{align}

As it is shown in \cite{Moti:2024b}, if a small box of such a fluid (consisting of spinninig point particles), in the presence of gravitational sources, is moved from point $P_1$ with coordinates $X$ to the point $P_2$ with coordinates $X+\Delta$,  then the conservation of the number current and the energy--momentum tensor is preserved, but the divergence of the entropy current feels a change, and we have
\begin{align}
      &  \Delta(\nabla_{\mu}\mathcal{N}^{\mu}_{\BM}) = 0 \\
      &  \Delta(\nabla_{\mu}\mathcal{T}^{\mu\nu}_{\BM})  = 0\\
      &  \Delta(\nabla_{\mu}\zeta^{\mu}_{\BM}) = - \dfrac{2k}{3} \left( R^{i}_{\ i\nu\alpha} + R^{\beta}_{\ \nu \beta\alpha} \right) \Delta^{\alpha} \mathcal{N}^{\nu}_{\BM}
\label{En-B-Eq}
\end{align}

The importance of the last equation shows up when we define the total entropy as
\begin{equation}
\varsigma_{\BM} = \int \dd[3]{x} \sqrt{g} \zeta^0_{\BM}
\end{equation}
and noting that by setting $\nabla_{\mu} \zeta^{\mu}_{\BM} = \cal{Q}$ and integrating
\begin{equation}
\int \dd[3]{x} \partial_{\mu} \left(\sqrt{g} \zeta^{\mu}_{\BM} \right)  = \int \dd[3]{x} \sqrt{g} \cal{Q}
\end{equation}
and on using equation \eqref{En-B-Eq}, the relation between entropy rate at the initial and final positions of the small box of fluid is given by \cite{Moti:2024b}
\begin{equation}
\Delta\dot{\varsigma}_{\BM} \equiv \dv{t} \bar{\varsigma}_{\BM} - \dv{t} \varsigma_{\BM} = -\dfrac{2k}{3} \left( R^{i}_{\ i\nu\alpha} + R^{i}_{\ \nu i\alpha} \right) \int \dd[3]{x} \sqrt{g} \Delta^{\alpha} \mathcal{N}^{\nu}_{\BM} 
\label{3rd m-balance eq.}
\end{equation} 
where the barred $\varsigma$ is the one at the final state. The equation is written in the Riemann normal coordinate system. This choice is justified as the points $P_1$ and $P_2$ are two neighboring points in the local (freely falling) frame.

In summary, the balance equations for the particle number four-flow and the energy--momentum tensor of the fluid in the box would not experience any changes under transportation from point $P_1$ to point $P_2$, but the entropy rate will do. From \eqref{3rd m-balance eq.} it is clear that this change is proportional to the Riemann curvature tensor, which means that transporting a small box filled by a fluid, causes the entropy rate to change by a term proportional to the Riemann tensor.

Thus, any deviation in the trajectory of the box, whether caused by placing the box in a curved spacetime or by the passage of a gravitational wave pulse, is expected to introduce a change in the rate of entropy. This means that the thermodynamics of the box does not return to its initial state, and thus exhibits some hysteresis behavior. 
Permanent changes in other macroscopic quantities produced by change in the gravitational field  can be obtained in a similar way. 

These results are in accordance with the statistical approach of previous subsection and shows the presence of Gravo-Thermo memory.

\section{An Ensemble of Spinning Extended Objects}
\label{Sec:III}
In the previous section we explored the mix of gravitational memory and thermal effects resulting to hysteresis effects for an ensemble of spinning point particles. But note that since we are talking about an ensemble of gravitating objects, a good candidate for it would be a cluster of celestial objects at thermal equilibrium. This could be a cluster of stars, a cluster of galaxies, and so on. For such systems, the approximation of point particles is not sufficient. Therefore it is a good idea to consider an ensemble of spinning extended objects.

In general relativity, systems would follow geodesics provided that their physical size is much smaller than the characteristic curvature length scale of the spacetime. When this condition is not satisfied, motion would be more complex, governed by Mathisson--Papapetrou--Dixon (MPD) equations. The trajectory followed by an extended object must instead be determined from the fundamental local conservation law of the energy--momentum tensor,\footnote{Note that the energy--momentum tensor appearing here carries the kinematic properties of a single extended object, rather than representing the statistical moments of a flow or an ensemble of particles.}
\begin{equation}
\nabla_{\mu}T^{\mu\nu}=0 \ .
\end{equation}
Tidal forces arising from spacetime curvature affect the dynamics of an extended object through local couplings between the curvature and the multipole moments of its energy--momentum tensor. These multipoles are defined through
\begin{equation}
I^{\mu_1 \ldots \mu_n \alpha\beta} = \int \delta x^{\mu_1} \ldots \delta x^{\mu_n} T^{\mu\nu} \sqrt{-g} \dd{\Sigma}
\end{equation}
where the determinant of spacetime metric $g_{\mu\nu}$ denoted with $g$, and the integration is performed over a   spacetime hypersurface $\Sigma$ at constant $t$. The $\delta x^{\mu} \equiv x^{\mu} - \mathcal{X}^{\mu}$ describes points about the worldline of a specified point (with coordinates $\mathcal{X}^{\mu}$) in the object.

Within this framework, a monopole particle is defined as an object for which at least one of the components of the zeroth moment integral of the energy--momentum tensor, $\int T^{\mu\nu} \sqrt{-g} \dd{\Sigma}$ is non-vanishing, while all higher multipole moments vanish. Any point particle is therefore described as a monopole. A monopole--dipole object, on the other hand, is the one with at least one non-vanishing dipole moment integral  $\int \delta x^{\rho}T^{\mu\nu} \sqrt{-g} \dd{\Sigma}$ , and vanishing higher moments, and so on.

In the rest frame of the object, the first and second moments, i.e.
\begin{align}
& P^{\mu} = \int T^{\mu 0} \sqrt{-g} \dd[3]{x} \ , \label{1st-Moment} \\
& S^{\mu\nu} = \int \left[ \delta x^{\mu} T^{0\nu}-\delta x^{\nu} T^{0\mu} \right] \sqrt{-g} \dd[3]{x} \ , \label{2nd-Moment}
\end{align}
are nothing but the linear momentum and the angular momentum about the point $\mathcal{X}^{\mu}$ (i.e. the spin of the extended object), respectively.

Then, the conservation equation, for these two leads to the following dynamical equations
\begin{align}
& \Ddv{P_\nu}{\tau} + \dfrac{1}{2} S^{\alpha\beta} R_{\alpha\beta\nu\mu} U^{\mu} = 0 \ , \\
& \Ddv{S^{\mu\nu}}{\tau} + U^{\mu}P^{\nu}-U^{\nu}P^{\mu} = 0 
\end{align}
where,  $U^{\mu}$ denotes the four-velocity of the reference point worldline $\mathcal{X}^{\mu}$ and $\tau$ is an affine parameter along the momentum four-vector $P^{\mu}$. This coupled system of ten equations, known as the Mathisson–Papapetrou equations, governs the dynamics of thirteen quantity: six independent components of the spin tensor $S^{\mu\nu}$, the four components of the momentum  $P^{\mu}$, and the three independent components of the four-velocity $U^{\mu}$, with the redundancy eliminated through the imposition of suitable physical constraints.

Physically use of the covariant center-of-mass frame of the extended body, ensures a unique and well-defined worldline for its effective description. This is defined through Tulczyjew--Dixon condition $P^{\mu}S_{\mu\nu} = 0$ which provides the covariant generalization of the rest frame condition of special relativity, $S^{0i}=0$.
Together with the conserved charges $m^2 = P_{\mu} P^{\mu}$ and $s^2 = \frac{1}{2}S_{\mu\nu}S^{\mu\nu}$, this condition closes the system and a complete set of equations governing the motion of an extended body in curved spacetime is provided \cite{Mathisson:1937,Papapetrou:1951, Corinaldesi:1951, Dixon:1970}
\begin{align}
& \Ddv{P_\nu}{\tau} + \dfrac{1}{2} S^{\alpha\beta} R_{\alpha\beta\nu\mu} U^{\mu} = 0 \ , \label{MP-Momentum}\\
& \Ddv{S_{\alpha\beta}}{\tau} + \dfrac{1}{m^2} \left(S_{\alpha\gamma} P_{\beta} + S_{\gamma\beta} P_{\alpha}   \right)  \Ddv{P^{\gamma}}{\tau}  = 0  \ , \label{MP-Spin} \\
& U^{\mu} = \dfrac{\bar{Z}}{m^2} \left( P^{\mu} +\dfrac{1}{2} ~ \dfrac{S^{\mu\nu} R_{\nu\gamma\alpha\beta}P^{\gamma}S^{\alpha\beta}}{m^2 + \dfrac{1}{4}R_{\alpha\beta\gamma\delta}S^{\alpha\beta}S^{\gamma\delta}}   \right) \ , \label{MP-Velocity}
\end{align}
in which $ \bar{Z} \equiv P_{\lambda} U^{\lambda} $ is a constant scalar quantity.

These are highly coupled, non-linear system of equations for which no analytic solution is known. Consequently, a variety of approximation schemes have been developed to study their solution, including perturbative expansions and semi-analytic treatments \cite{Tod:1976, Mashhoon:2006, Singh:2008, Mohseni:2000}. For our purpose, a combination of perturbative and numerical methods provides an effective framework for tracking finite-size effects and their associated memory imprints along the worldline.

\begin{figure}
\centering
\begin{subfigure}{.48\textwidth}
    \centering
    \includegraphics[width=.99\linewidth]{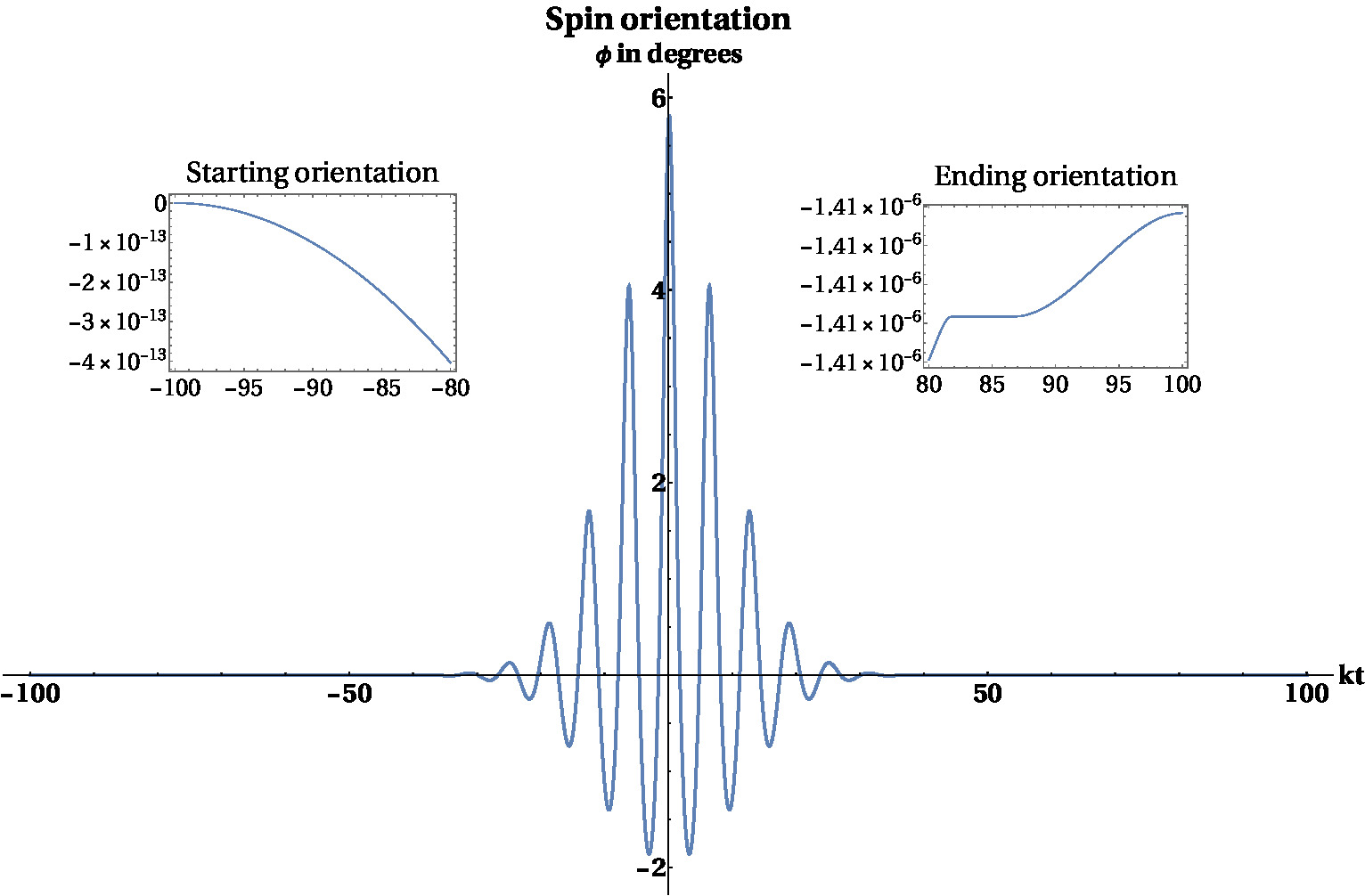}
    \caption{Typical small total change in the azimuthal angle.} 
    \label{phi-HpTHc-10}
\end{subfigure}
\begin{subfigure}{.48\textwidth}
    \centering
    \includegraphics[width=.99\linewidth]{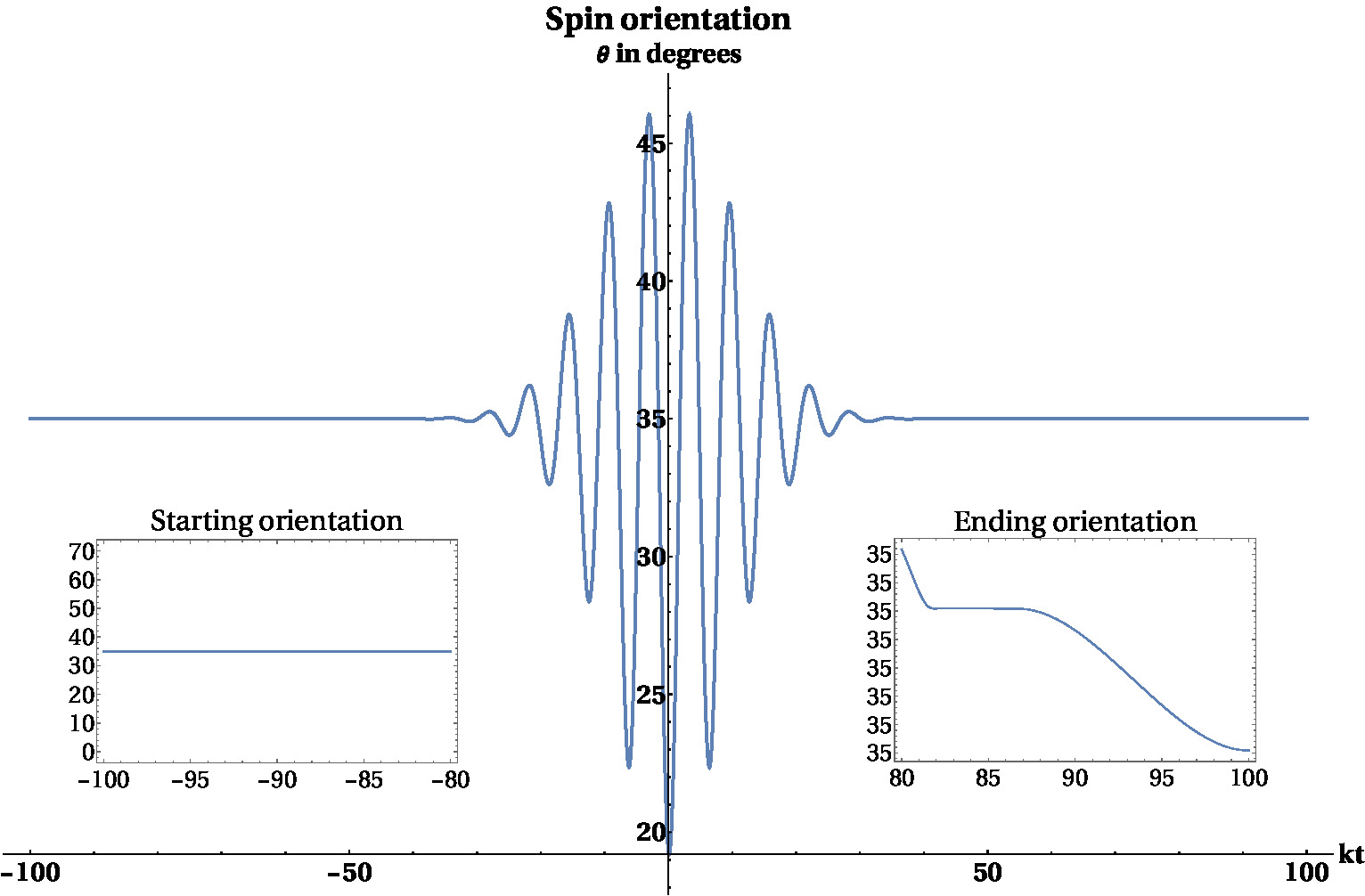}
    \caption{Near-vanishing total change in zenith angle.}
    \label{theta-HpTHc-10}
\end{subfigure}
\begin{subfigure}{.48\textwidth}
    \centering
    \includegraphics[width=.99\linewidth]{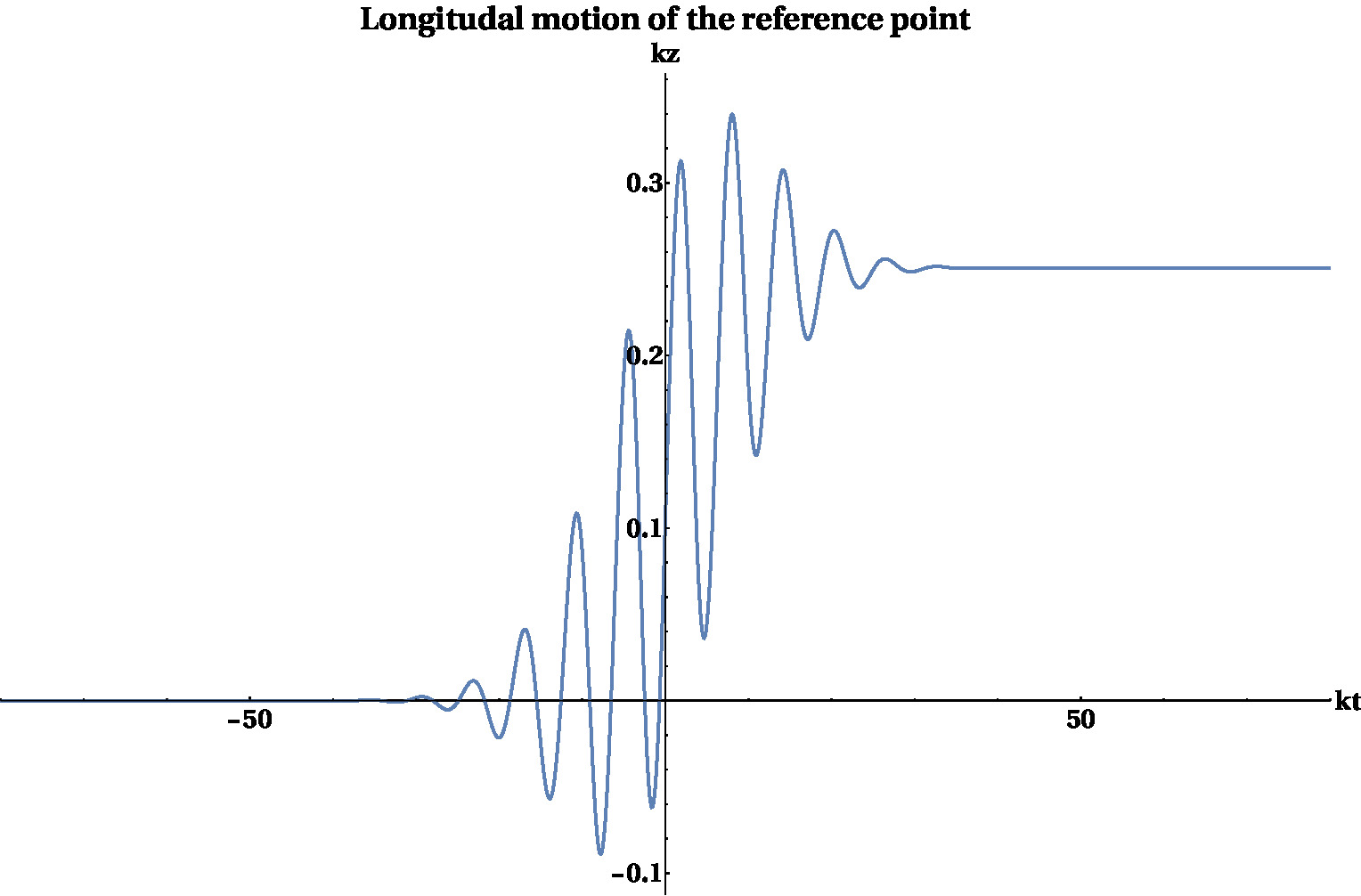}
    \caption{Typical longitudal motion of the reference point.}
    \label{LonMo-HpTHc-10}
\end{subfigure}
\begin{subfigure}{.45\textwidth}
    \centering
    \includegraphics[width=.99\linewidth]{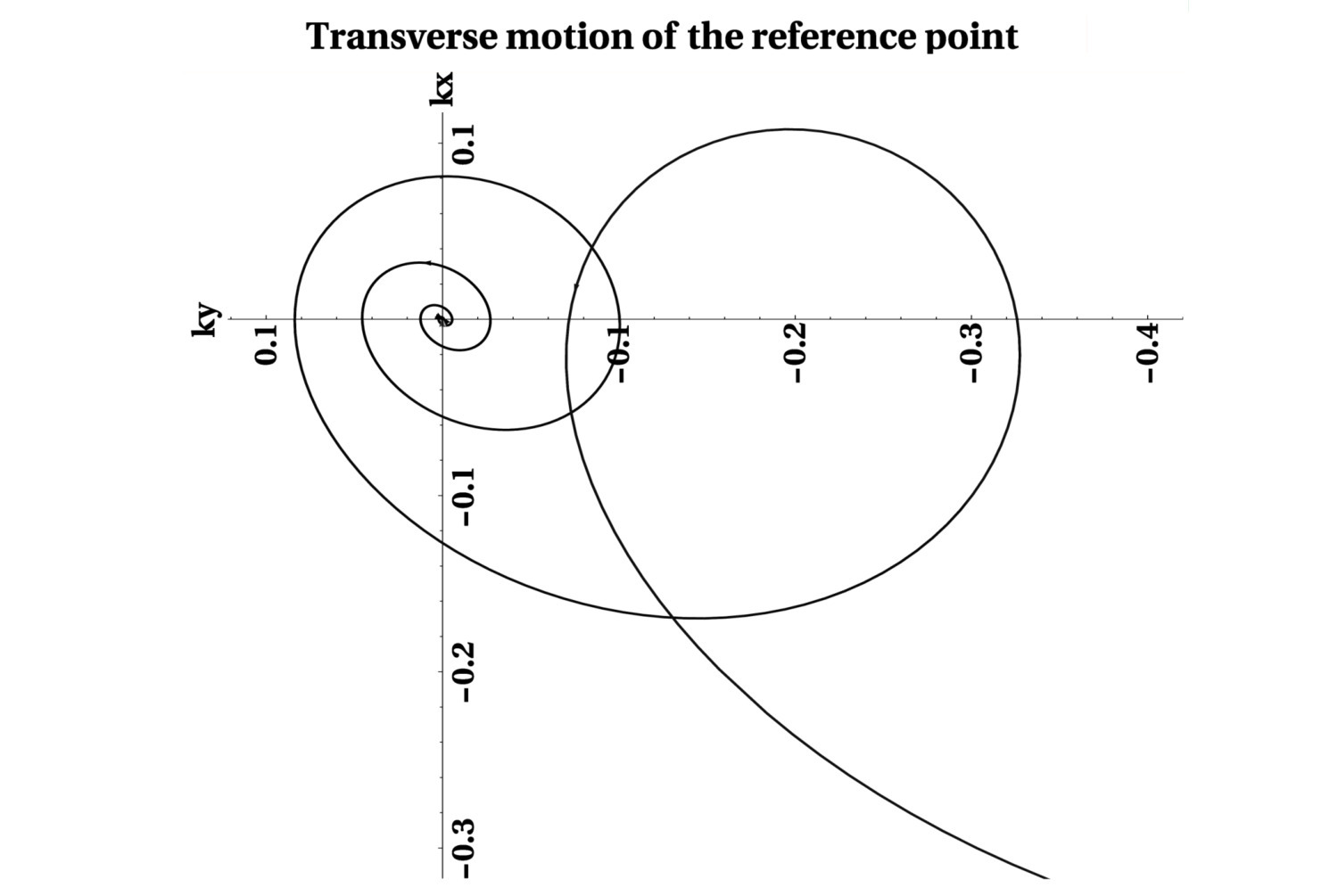}
    \caption{Typical transverse motion of the reference point.}
    \label{TraMo-HpTHc-10}
\end{subfigure}
\caption{Typical Gravo-Thermo memory in the motion of an extended spinning object \cite{Moti:2025}.}
\label{SpiExtMo}
\end{figure}

To see the effect as it is shown in \cite{Moti:2025} consider a flat spacetime perturbed by gravitational wave, defined with
\begin{equation}
g_{\mu\nu}^{\text{(Cartesian)}}=
\begin{pmatrix}
1 & 0 & 0 & 0 \\
0 & -1- \epsilon h_{+}(x^{\mu})  & \epsilon h_{\times}(x^{\mu})  & 0 \\
0 & \epsilon h_{\times}(x^{\mu})  & -1 + \epsilon h_{+}(x^{\mu})  & 0 \\
0 & 0 & 0 & -1
\end{pmatrix}
\end{equation}
and expand the dynamical properties as 
\begin{align}
& P^{\mu} = P^{\mu}_{(0)} + \epsilon P^{\mu}_{(1)} \ , \\
& S^{\mu\nu} = S^{\mu\nu}_{(0)} + \epsilon S^{\mu\nu}_{(1)} \ , \\
& U^{\mu} = U^{\mu}_{(0)} + \epsilon U^{\mu}_{(1)} 
\end{align}
where $P^{\mu}_{(1)}$, $S^{\mu\nu}_{(1)}$, and $ U^{\mu}_{(1)} $ denote the first order contributions to the momentum, spin, and velocity, respectively and $\epsilon$ is the small parameter of perturbation.
For properly chosen initial conditions, it was shown in \cite{Moti:2025} that the passage of a gravitational pulse induces permanent changes in the transverse motion, longitudinal motion, and spin orientations of the extended object. Figure \ref{SpiExtMo} shows typical motion as a gravitational wave pulse passes the object. Refer to \cite{Moti:2025} for details.

As it is expected, during the passage of the pulse, the dynamics exhibit a transient oscillatory phase, followed by a non-oscillatory sector that persists after the pulse has passed. It is this latter sector that encodes the memory effect. 

A notable feature, consistent with the point particle case, is that one of the orientation degrees of freedom remains unaffected by the memory effect at this first order of  perturbation, figure \ref{theta-HpTHc-10}. This is in complete agreement with what is discussed in \ref{App. A}.

\subsection{Statistical Description}
\label{SecIII:Thermo}

As for the case with point particles, thermal effects can be added using the statistical concepts and the obtained numerical solution. As it is shown in \cite{Moti:2025} we will encounter with a partition function like
\begin{equation}
\mathcal{Z}(T) \sim \int e^{-\beta H_{\text{int}}} \dd{\Omega}
\end{equation}
for a collection of $N$ randomly oriented spinning objects, where the orientation of the $i$-th object is described by the angles $(\theta_i,\phi_i)$ for $i=1,\ldots, N$. Assuming that there is a magnetic field in the $x$-direction interacting with the magnetic moments of the objects, the Hamiltonian would be $H = H_0 + H_{\text{int}}(\theta_i,\phi_i)$.

The memory sector of the gravitational wave pulse, reorients each component to $(\theta_i+\Delta\theta_i, \phi_i+\Delta\phi_i)$. Consequently, the interaction Hamiltonian changes to $H_{\text{int}}(\theta_i+\Delta\theta_i, \phi_i+\Delta\phi_i)$. This shift from the initial configuration results in a change in the partition function, $\Delta\mathcal{Z}$.

The values of $(\Delta\theta_i, \Delta\phi_i)$ depend on the system dynamics where managed by the Mathisson--Papapetrou equations.
As a result, the effect of the initial orientations on the final distribution is encoded through the dependence of $\Delta\phi_i$ on $\theta_i$, as confirmed by the numerical results.

These considerations allow us to simplify the analysis by replacing the integration over $\theta_i$ with a discrete averaging. Therefore
\begin{equation}
 \mathcal{Z}(T) \sim \dfrac{\pi}{N} \sum_{i=1}^N \mathcal{Z}_i(T,\theta_i) \ ,
 \label{Z}
\end{equation}
where
\begin{equation}
\mathcal{Z}_i(T,\theta_i) = \int e^{-\beta H_{\text{int}}(\theta_i, \phi+\Delta\phi_i(\theta_i))} \sin{\theta_i} \dd{\phi} \ .
\label{Z-theta}
\end{equation}
To evaluate the partition function \cite{Moti:2025}, one can assume a random distribution of $\theta_i$ values. This enables us to evaluate $\Delta\mathcal{Z}$, i.e. the change in the partition function due to the passage of the gravitational wave pulse. This represents the gravitational memory encoded into the partition function and thus in the thermodynamic quantities.

Like the case of point particles, assuming an ensemble of extended objects with magnetic moments placed in a magnetic field,  the interaction Hamiltonian is given by $ H_{\text{int}} \sim \mu \vec{B}\vdot \vec{S}$. Thus
\begin{equation}
H_{\text{int}} \sim \mu B S_{\perp} \cos{\phi}
\end{equation} 
where $S_{\perp} = \abs*{\vec{S}} \sin{\theta_i} $  which is the non-vanishing spin contribution to the Hamiltonian. Using this, we have 
\begin{equation}
\mathcal{Z}_i(T,\theta_i) = \int e^{- \frac{T_0}{T} \sin{\theta_i} \cos{(\phi+\Delta\phi_i(\theta_i))}} \sin\theta_i \dd{\phi}
\label{Z_i}
\end{equation}
where $T_0=\mu B \abs*{\vec{S}}/k_B$ is a constant with the unit of temperature, playing the role of some characteristic temperature for the system.

Evaluating $\mathcal{Z}_i(T,\theta_i)$ using the numerical values of $\Delta\phi_i(\theta_i)$ is straightforward and can be found in details in \cite{Moti:2025}. This means that the partition function undergoes a change $\Delta\mathcal{Z}$ after redistribution of spins because of interaction with the incoming wave which leads to deviation of statistical quantities like entropy and internal energy from their initial value.  This behavior is qualitatively similar to that of an ensemble of point particles. However, for appropriately chosen parameters the ensemble of extended objects exhibits a resonance in its energy response at some specific temperature \cite{Moti:2025}. This resonance indicates that the ensemble can efficiently absorb energy from the gravitational wave pulse at a specific temperature (which depends on the parameters).

\subsection{Kinetic Theoretical Description}
\label{SecIII:Kinetic}

We are now in the position to present a kinetic theoretical description of a flow of extended spinning objects. To do so, we need to generalize the Boltzmann equation for a flow of such objects in the presence of gravity, and obtain the governing equations of different moments of multipole expansion of the Boltzmann equation. Then we need to compare these balance equations for a box of the fluid when moved from a point to its neighboring points.

A key new feature is the explicit inclusion of the spin moment  $S^{\mu\nu}$ as an additional degree of freedom in the distribution function $f(x^{\mu}, P^{\mu}, S^{\mu\nu})$. The presence of spin significantly enriches the underlying structure of the theory, leading to a more general and considerably more intricate dynamical framework. In order to make the text easily readable, the intermediate and lengthy steps of the calculations presented here, are provided in \ref{App. B}.

For $f \equiv f(x^{\mu}, P^{\mu}, S^{\mu\nu})$ the generalized Boltzmann equation is
\begin{equation}
\dv{f}{\tau} = \pdv{f}{x^{\mu}} \dv{x^{\mu}}{\tau} + \pdv{f}{P^{\mu}} \dv{P^{\mu}}{\tau} + \pdv{f}{S^{\mu\nu}} \dv{S^{\mu\nu}}{\tau} = Q 
\label{Ex-Boltzman}
\end{equation}
where $Q \equiv Q(f, F,\Omega, P)$, as before.

Using definitions of the directional covariant derivatives $\LDdv{P^\alpha}{\tau}$ and $ \LDdv{S^{\alpha\beta}}{\tau}$ together with the equation \eqref{MP-Velocity}, the effect of gravity enters the Boltzmann equation \eqref{Ex-Boltzman} through
\begin{multline}
 \dfrac{\bar{Z}}{m^2} \pdv{f}{x^{\mu}} \left( P^{\mu} + \tilde{\Pi}^{\mu}   \right) +
\pdv{f}{P^{\mu}} \left( \Ddv{P^{\mu}}{\tau} - \dfrac{\bar{Z}}{m^2}  \Gamma^{\mu}_{\alpha\beta} \left( P^{\alpha} + \tilde{\Pi}^{\alpha}   \right) P^{\beta} \right) + \\
\pdv{f}{S^{\mu\nu}} \left(  \Ddv{S^{\mu\nu}}{\tau} - \dfrac{\bar{Z}}{m^2}  \left( \Gamma^{\mu}_{\alpha\beta}  S^{\beta\nu} + \Gamma^{\nu}_{\alpha\beta}  S^{\mu\beta}\right) \left( P^{\alpha} + \tilde{\Pi}^{\alpha}   \right) \right) = Q
\label{3rd-Ed}
\end{multline}
where $\tilde{\Pi}^{\mu} = \dfrac{1}{2} ~ \left ( \dfrac{S^{\mu\nu} R_{\nu\sigma\alpha\beta}P^{\sigma}S^{\alpha\beta}}{m^2 + \dfrac{1}{4}R_{\alpha\beta\gamma\delta}S^{\alpha\beta}S^{\gamma\delta}} \right )$.

Neglecting the higher orders of the spacetime perturbations, one arrives at
\begin{multline}
\pdv{f}{x^{\mu}} P^{\mu}   -
\pdv{f}{P^{\mu}} \Gamma^{\mu}_{\alpha\beta}  P^{\alpha} P^{\beta}  - \pdv{f}{S^{\mu\nu}} \left( \Gamma^{\mu}_{\alpha\beta}  S^{\nu\beta} + \Gamma^{\nu}_{\alpha\beta}  S^{\mu\beta}\right)  P^{\alpha} = \\
-\pdv{f}{x^{\mu}} \tilde{\Pi}^{\mu} -  \dfrac{m^2}{\bar{Z}} \pdv{f}{P^{\mu}} \Ddv{P^{\mu}}{\tau} - \dfrac{m^2}{\bar{Z}}  \pdv{f}{S^{\mu\nu}}  \Ddv{S^{\mu\nu}}{\tau}  + Q
\label{4th-Ed}
\end{multline}

Like the case for point particles presented in \cite{Moti:2025}, the balance equations for the essential moments are obtained by averaged Boltzmann equation weighted by an appropriate function $\psi$. Integrating the Boltzmann equation multiplied by $\psi$ over the extended phase-space measure $(g/P_0)  \dd{\chi}_s  \equiv  (g/P_0)  \dd[3]{P} \dd[4]{x} \dd{\Omega_s} $, yields
\begin{multline}
\int \psi \left[ \pdv{f}{x^{\mu}} P^{\mu}   -
\pdv{f}{P^{i}} \Gamma^{i}_{\alpha\beta}  P^{\alpha} P^{\beta}  - \pdv{f}{S^{\mu\nu}} \left( \Gamma^{\mu}_{\alpha\beta}  S^{\nu\beta} + \Gamma^{\nu}_{\alpha\beta}  S^{\mu\beta}\right)  P^{\alpha} \right] \dfrac{g}{P_0} \dd{\chi}_s \\
= - \int \psi \left[ \pdv{f}{x^{\mu}} \tilde{\Pi}^{\mu} +  \dfrac{m^2}{\bar{Z}} \pdv{f}{P^{i}} \Ddv{P^{i}}{\tau} + \dfrac{m^2}{\bar{Z}}  \pdv{f}{S^{\mu\nu}}  \Ddv{S^{\mu\nu}}{\tau} \right]   \dfrac{g}{P_0}  \dd{\chi}_s + \int \psi \mathcal{Q} \dfrac{g}{P_0} \dd{\chi}_s
\label{MainEq}
\end{multline}
Note that the measure $\dd{\Omega_s}=[\dd[4]{S}] $ represents the spin-space volume element and includes integration over all possible spin magnitudes and orientations subjected to all constraints on it.

Use of the Liouville Theorem besides the MPD dynamics, helps us to have a more compact form
\begin{align}
& \int \sqrt{-g} \dd[4]{x} \left[ \nabla_{\mu} \int \left( f \psi  P^{\mu} \sqrt{-g} \dfrac{\dd[3]{P} }{P_0}  \dd{\Omega}_s \right)  \right. \nonumber \\
& \left. -  \int  f \left( \pdv{\psi}{x^{\mu}}   P^{\mu}   - \pdv{\psi}{P^{i}}  \Gamma^{i}_{\mu\nu} P^{\mu} P^{\nu}  - \pdv{\psi}{S^{\mu\nu}}   \left( \Gamma^{\mu}_{\alpha\beta}  S^{\nu\beta} + \Gamma^{\nu}_{\alpha\beta}  S^{\mu\beta}\right) P^{\alpha} \right) \sqrt{-g} \dfrac{\dd[3]P}{P_0}  \dd{\Omega}_s \right] \nonumber \\
& = \int \sqrt{-g} \dd[4]{x} \left( \int \mathcal{D}_{\text{MPD}} \sqrt{-g} \dfrac{\dd[3]{P}}{P_0} \dd{\Omega}_s  + \int \psi \mathcal{Q} \sqrt{-g} \dfrac{\dd[3]{P}}{P_0} \dd{\Omega}_s \right)
\label{MainEq2}
\end{align}
where the term
\begin{equation}
\mathcal{D}_{\text{MPD}} \equiv -\psi \left[ \pdv{f}{x^{\mu}} \tilde{\Pi}^{\mu} +  \dfrac{m^2}{\bar{Z}} \left( \pdv{f}{P^{\mu}} g^{\mu\kappa} - \dfrac{1}{m^2}  \pdv{f}{S^{\mu\nu}}  \left(  S^{\mu\kappa} P^{\nu} + S^{\kappa\nu} P^{\mu} \right)    \right) \Ddv{\tilde{P}_{\kappa}}{\tau}   \right]  
\label{D-MPD}
\end{equation}
carries the effects of MPD dynamics.
The first term in the brackets on the $l.h.s$ corresponds exactly to the balance equations.

To investigate the effect of hysteresis on the conservation of the essential moments, one should compare this equation at two distinct spacetime points. Consider a finite volume box containing a fluid of extended spinning objects. We are interested in seeing how transporting this box from the point $X$ to $X+\Delta$ affects the balance equations. To see this, one has to evaluate the difference between the above equation at these two spacetime locations, namely
\begin{align}
& \left[ \int \left( \bar{f} \bar{\psi}  P^{\mu} \sqrt{-g} \dfrac{\dd[3]{P} }{P_0}  \dd{\Omega}_s \right)  \right]_{;\mu} -\left[ \int \left( f \psi P^{\mu} \sqrt{-g} \dfrac{\dd[3]{P} }{P_0}  \dd{\Omega}_s \right)  \right]_{;\mu} \nonumber \\
& =  \int  \left( \bar{f}\pdv{\bar{\psi}}{x^{\mu}}  - f  \pdv{\psi}{x^{\mu}} \right) P^{\mu}\sqrt{-g} \dfrac{\dd[3]P}{P_0}  \dd{\Omega}_s  - 
\int \left( \bar{f}  \pdv{\bar{\psi}}{P^{i}}  \bar{\Gamma}^{i}_{\mu\nu} - f \pdv{\psi}{P^{i}}  \Gamma^{i}_{\mu\nu}  \right) \sqrt{-g} P^{\mu} P^{\nu}  \dfrac{\dd[3]P}{P_0}  \dd{\Omega}_s \nonumber \\
& - \int \left[\bar{f}  \pdv{\bar{\psi}}{S^{\mu\nu}}\left( \bar{\Gamma}^{\mu}_{\alpha\beta}  S^{\nu\beta} + \bar{\Gamma}^{\nu}_{\alpha\beta} S^{\mu\beta} \right) - f \pdv{\psi}{S^{\mu\nu}} \left( \Gamma^{\mu}_{\alpha\beta}  S^{\nu\beta} + \Gamma^{\nu}_{\alpha\beta}  S^{\mu\beta} \right) \right]  P^{\alpha}  \sqrt{-g} \dfrac{\dd[3]P}{P_0}  \dd{\Omega}_s  \nonumber \\
& + \int \left(\bar{\mathcal{D}}_{\text{MPD}}-\mathcal{D}_{\text{MPD}} \right) \sqrt{-g} \dfrac{\dd[3]{P}}{P_0} \dd{\Omega}_s + \int \left(\bar{\psi} \bar{\mathcal{Q}}-\psi \mathcal{Q} \right) \sqrt{-g} \dfrac{\dd[3]{P}}{P_0} \dd{\Omega}_s
\label{De-Main}
\end{align}
where the barred quantities are the ones evaluated at $X+\Delta$.

For linear perturbation $\bar{f} \equiv \bar{f}(X+\Delta) = f(X) + \delta{f}(X)$ and $ \bar{\Gamma}^{\gamma}_{\alpha\beta} \equiv \bar{\Gamma}^{\gamma}_{\alpha\beta}(X+\Delta) = \Gamma^{\gamma}_{\alpha\beta}(X) + \Delta^{\rho} \partial_{\rho} \Gamma^{\gamma}_{\alpha\beta}(X)$, we can examine the behavior of the three main kinetic theory moments, i.e,
\begin{align}
      & \mathcal{N}^{\mu}_{\BM_{s}} = c \displaystyle\int f  P^{\mu} \sqrt{-g} \dfrac{\dd[3]{P}}{P_0} \dd{\Omega}_s  \ , \label{N-MPD-Mom} \\
      & \mathcal{T}^{\mu\nu}_{\BM_{s}} = c \displaystyle\int  f P^{\mu} P^{\nu} \sqrt{-g} \dfrac{\dd[3]{P}}{P_0} \dd{\Omega}_s   \ , \label{T-MPD-Mom} \\
      & \zeta^{\mu}_{\BM_{s}} = -ck  \displaystyle\int  f P^{\mu}\ \ln{\left (\dfrac{f h^3}{e g_s}\right )}\     \sqrt{-g} \dfrac{\dd[3]{P}}{P_0} \dd{\Omega}_s  \ . \label{Ent-MPD-Mom}
\end{align}
Note that the subscript `$s$' is introduced to distinguish these three quantities, which account for the spin degrees of freedom of the extended object, from their corresponding point particle counterparts, i.e. equations \eqref{PN-Mom} to \eqref{Ent-Mom}.

Now, let's investigate the change in the balance equations due to different choices of $\psi$:
\begin{itemize}
\item  {\bf Number current conservation}: For $(\psi,\bar{\psi}) = (c,c)$, together with defining
\begin{equation}
 \mathcal{N}^{\mu}_{\BM_{s}} = c \int f P^{\mu}  \sqrt{-g} \dfrac{\dd[3]{P}}{P_0} \dd{\Omega}_s,  \qquad  \bar{\mathcal{N}}^{\mu}_{\BM_{s}} = c \int\bar{f}  P^{\mu} \sqrt{-g} \dfrac{\dd[3]{P}}{P_0} \dd{\Omega}_s
\label{psi-N}
\end{equation} 
the equation \eqref{De-Main} leads to
\begin{equation}
\Delta(\nabla_{\mu}\mathcal{N}^{\mu}_{\BM_{s}}) \equiv \nabla_{\mu}\bar{\mathcal{N}}^{\mu}_{\BM_{s}} -\nabla_{\mu}\mathcal{N}^{\mu}_{\BM_{s}}  = \mathcal{O} (2) \ .
\label{51}
\end{equation}
This vanishing of this up to the first order of perturbation implies this trivial fact that gravity does not induce particle creation.  Therefore, the number current is unaffected by the spin degrees of freedom of the objects and remains conserved, as in the case of an ensemble of point particles. Clearly this is true at any order of perturbation, and thus we can put the $r.h.s$ of equation (\ref{51}) identically equal to zero.

\item {\bf Energy--momentum conservation}: Choosing the set of four-vectors  $(\psi^{\nu}, \bar{\psi}^{\nu}) = (cP^{\nu},cP^{\nu})$, and using the definition of the energy--momentum tensor
\begin{equation}
 \mathcal{T}^{\mu\nu}_{\BM_{s}} = c \int f P^{\mu} P^{\nu}  \sqrt{-g} \dfrac{\dd[3]{P}}{P_0} \dd{\Omega}_s, \qquad  \bar{\mathcal{T}}^{\mu\nu}_{\BM_{s}} = c \int \bar{f} P^{\mu} P^{\nu}  \sqrt{-g} \dfrac{\dd[3]{P}}{P_0} \dd{\Omega}_s\
\label{psi-T}
\end{equation}
we can explore the balance equation of the energy--momentum tensor.
The covariant form of \eqref{App-Cov-Bolt-Ave} for $\psi^{\nu}$, leads to
\begin{align}
\Delta(\nabla_{\mu}\mathcal{T}^{\mu\nu}_{\BM_{s}}) \equiv \nabla_{\mu}\bar{\mathcal{T}}^{\mu\nu}_{\BM_{s}} -\nabla_{\mu}\mathcal{T}^{\mu\nu}_{\BM_{s}} & = \mathcal{O} (2) \ .
\end{align}
Consequently, the dynamics of the extended objects have no effect on the system at this order. However, such effects may emerge at higher orders in the perturbative expansion.

\item {\bf Entropy rate equation}: At last if we choose $(\psi, \bar{\psi}) = \left(- ck \ln{\left(\dfrac{f h^3}{e g_s}\right)}, - ck \ln{\left(\dfrac{\bar{f} h^3}{e g_s}\right)} \right)$ and use the entropy-flux definition at two points, i.e.
\begin{equation}
\zeta^{\mu}_{\BM_{s}} = - ck  \int f  \ln{\left(\dfrac{f h^3}{e g_s}\right)}    P^{\mu} \sqrt{g} \dfrac{ \dd[3]{p} }{p_0}\dd{\Omega}_s, \qquad \bar{\zeta}^{\mu}_{\BM_{s}} = - ck  \int  \bar{f}   \ln{\left(\dfrac{\bar{f} h^3}{e g_s}\right)} P^{\mu} \sqrt{g} \dfrac{\dd[3]{p}}{p_0} \dd{\Omega}_s 
\end{equation}
after some calculations we would get
\begin{equation}
\Delta (\nabla_{\mu} \zeta^{\mu}_{\BM_{s}}) = \nabla_{\mu} \bar{\zeta}^{\mu}_{\BM_{s}}-\nabla_{\mu} \zeta^{\mu}_{\BM_{s}} - 
\dfrac{2k}{3} \left( R^{i}_{\ i\nu\alpha} + R^{i}_{\ \nu i \alpha} \right) \Delta^{\alpha} \mathcal{N}^{\rho}_{\BM_{s}} > 0
\end{equation}
The precence of the last term changes the balance equation $\nabla_{\mu} \zeta^{\mu}_{\BM_{s}} > 0 $, as an effect of the memory sector. Other terms are ignored with respect to this leading term.
\end{itemize}

Besides these three essential moments, two further moments are needed to characterize the rotational degrees of freedom of the spinning extended objects constituting the fluid,
\begin{align}
      & \Sigma^{\mu\nu \tau}_{\BM_{s}} =  \displaystyle\int  f  S^{\mu\nu} P^{\tau}\sqrt{-g} \dfrac{\dd[3]{P}}{P_0} \dd{\Omega}_s  \label{S-MPD-Mom} \\
      & L^{\mu\nu\tau}_{\BM_{s}} =   \displaystyle\int f \left( X^{\mu}P^{\nu} - X^{\nu} P^{\mu}    \right) P^{\tau}  \sqrt{-g} \dfrac{\dd[3]{P}}{P_0} \dd{\Omega}_s  \label{L-MPD-Mom}
\end{align}
which are spin and orbital angular momentum current tensors, respectively.

To derive the change in the spin-current balance equation, we use
\begin{align}
& \psi^{\mu\nu} = cS^{\mu\nu}, \quad \quad 
 \Sigma^{\mu\nu\tau}_{\BM_{s}} = c \int f S^{\mu\nu} P^{\tau}  \sqrt{-g} \dfrac{\dd[3]{P}}{P_0} \dd{\Omega}_s \nonumber \\
& \bar{\psi}^{\mu\nu} = c\bar{S}^{\mu\nu} , \quad \quad  \bar{\Sigma}^{\mu\nu\tau}_{\BM_{s}} = c \int \bar{f} S^{\mu\nu} P^{\tau}  \sqrt{-g} \dfrac{\dd[3]{P}}{P_0} \dd{\Omega}_s \ .
\end{align} 
Following the same process as for energy--momentum tensor, we would have (see \ref{App. B})
\begin{equation}
\Delta \left( \nabla_{\tau}\Sigma^{\mu\nu\tau}_{\BM_s} \right) =   
\dfrac{4}{3} \Delta^{\rho} R^{\left[ \mu \right.}_{\ (\tau\alpha)\rho} \Sigma_{\BM_s}^{\left. \nu \right] \alpha\tau}
\label{Spin-Cons}
\end{equation}
The change in the conservation of the spin current is expected because of the interaction of the fluid with the gravitational field. Also note that the higher order moments are neglected in the MPD dynamics. To recover the conservation of total angular momentum, the balance equation for the orbital angular momentum must be considered alongside with that of the spin angular momentum. 

On the other hand for orbital angular momentum, we use the sets
\begin{align}
& \psi^{\mu\nu} = c  \left( X^{\mu}P^{\nu} - X^{\nu} P^{\mu}    \right), \qquad L^{\mu\nu\tau}_{\BM_{s}} =  c \int f \left( X^{\mu}P^{\nu} - X^{\nu} P^{\mu}    \right) P^{\tau}  \sqrt{-g} \dfrac{\dd[3]{P}}{P_0} \dd{\Omega}_s \nonumber \\
& \bar{\psi}^{\mu\nu} = c  \left( \bar{X}^{\mu}P^{\nu} - \bar{X}^{\nu} P^{\mu}    \right), \qquad \bar{L}^{\mu\nu\tau}_{\BM_{s}} =  c \int  \bar{f} \left( \bar{X}^{\mu}P^{\nu} - \bar{X}^{\nu} P^{\mu}    \right) P^{\tau} \sqrt{-g} \dfrac{\dd[3]{P}}{P_0} \dd{\Omega}_s
\end{align}
besides the expansion of $\bar{X}^{\alpha}$ and $ \bar{\Gamma}^{\gamma}_{\alpha\beta}$ one would have
\begin{equation}
\Delta(\nabla_{\tau} L^{\mu\nu\tau}_{\BM_s}) = \dfrac{4}{3} \Delta^{\rho}   R^{\left[ \mu \right. }_{\ (\tau\alpha)\rho} L_{\BM_s}^{\left.  \nu \right] \alpha\tau} 
\end{equation}

In the case where no angular momentum is delivered to the fluid by the gravitational interaction (e.g. by absorbing gravitational waves and their helicity), no change is present in the total angular momentum conservation
\begin{equation}
\Delta(\nabla_{\tau} J^{\mu\nu\tau}_{\BM_s}) = 0
\end{equation} 
where $J^{\mu\nu\tau}_{\BM_s} = L^{\mu\nu\tau}_{\BM_s} + \Sigma^{\mu\nu\tau}_{\BM_s} $. The condition for this is to have
\begin{equation}
\Delta^{\rho}   R^{\left[ \mu \right. }_{\ (\tau\alpha)\rho} J_{\BM_s}^{\left.  \nu \right] \alpha\tau}=0
\end{equation}

\section{Concluding Remarks}
\label{Sec:IV}

The intrinsic non-linearity of the Einstein equations can give rise to hysteresis effects in systems which have interaction with gravity. These effects may be manifested observationally as permanent displacements, relative velocities, spin precession, and as the related memory phenomena. For an ensemble of objects, such irreversible changes can modify the system partition function and thereby admit an effective thermodynamic description with observable consequences.

Here, we first brought the  Gravo-Thermo memory developed in \cite{Moti:2024a, Moti:2024b, Moti:2025} into a concise framework. In those studies, the concept of  Gravo-Thermo memory is introduced for point particles and then extended to the case of extended spinning objects, partially and from different aspects. 

Based on the central idea of  Gravo-Thermo memory for both an ensemble of  point particles and extended spinning objects, a versatile kinetic theoretical model is presented within which the change in the balance equations due to change in gravity (whether produced by moving a box of such fluid, or exposing the box by a gravitational wave) is examined.
We saw that at lowest order of perturbation, the conservation of particle number, energy--momentum and total angular momentum is preserved, while the entropy rate will be changed. Therefore the thermodynamics of a general system of spinning point-like or extended objects carry the imprint of the history of the gravity dynamics.  

Although the present model is formulated in the context of gyroscope memory, the underlying idea is more general. Any interaction between gravity and an ensemble of objects can induce permanent, history dependent modifications of the ensemble, thereby distinguishing its current state from its past configurations. Such irreversible imprints can be captured through changes in the thermodynamics of the system and consequently can lead to observable effects.

\appendix
\renewcommand{\thesection}{Appendix \Alph{section}}
\renewcommand{\theequation}{\Alph{section}.\arabic{equation}}

\section{Some Physical Assumptions} \label{App. A}
\setcounter{equation}{0}

Here, we examine the physical assumptions underlying the derivation of the analytical expression in equation \eqref{PF}. These assumptions are in exact consistency  with the numerical solutions of the extended object model.

\begin{figure}
\begin{centering}
\begin{subfigure}{.4\textwidth}
	\centering
	\includegraphics[width=.99\linewidth]{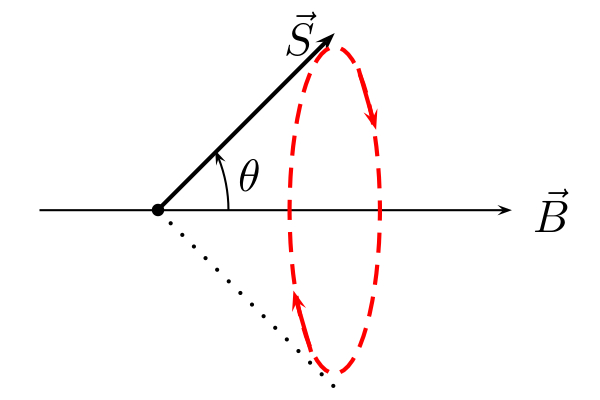}
	\caption{Orientation of a classical spinning object}
	\label{spinning-particle}
\end{subfigure}
\hglue1.5cm
\begin{subfigure}{.3\textwidth}
    \centering
	\includegraphics[width=.99\linewidth]{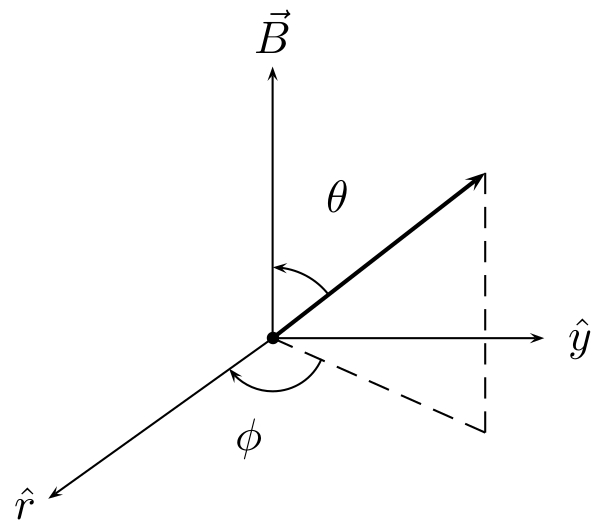}
	\caption{Spin orientation relative to the magnetic field $\vec{B}$}
	\label{orientation}
\end{subfigure}
\caption{Spin and magnetic field relative configuration.}
\label{SpiFieRelConf}
\end{centering}
\end{figure}

A single classical spinning object in a magnetic field (forming an angle $\theta$ with it), would have precession about the field, as in figure \ref{spinning-particle}.
As discussed in the text, the effect of the gravitational wave is to change the angles $(\theta,\phi)$. In fact, the spin can be decomposed to two parts
\begin{align}
& \vec{S}_0 = S_r \hat{r} + S_x \hat{B} + S_y \hat{y} \\
& \Delta \vec{S} = \Delta\Phi (S_y \hat{B} - S_x \hat{y})
\end{align}
where $ (S_x, S_y) = S_0(\cos\theta, \sin\theta \sin\phi) $ and $\vec{S} $ is the length of projection of $\vec{S}_0$ on $(\hat{B},\hat{y})$-plane, figure \ref{orientation}.

Looking at an ensemble of spinning particles \eqref{NonP-PF}, a correction of order $(\Delta\Phi)^2$ appears as a result of non-oscillating sector of gravitational wave memory. But, this is not realistic to think of magnetic matters as an infinite ensemble of non-interacting spinning objects. There are many phenomena to consider. 

\begin{figure}[h!]
\centering
\includegraphics[width=.5\linewidth]{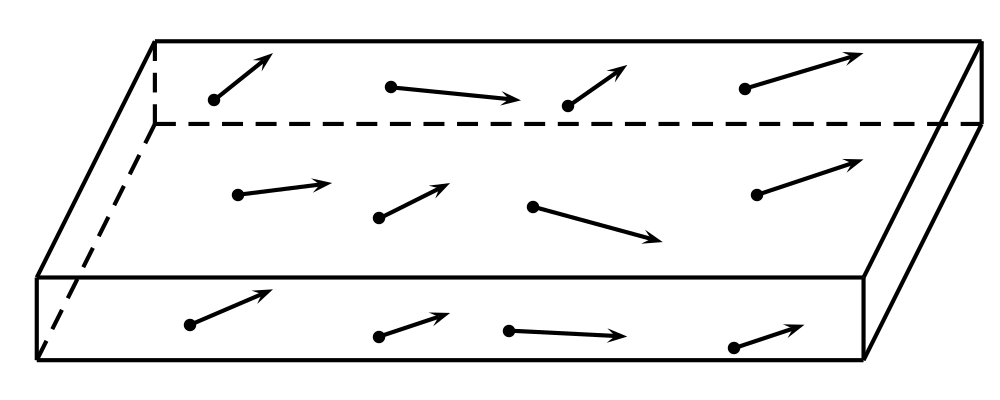} 
\caption{Volume finiteness effects}
\label{friction}
\end{figure}

For example, the weak spin--spin interaction between neighboring spins forces them to have more in-phase precession.
Moreover, there are other interactions that can be modeled as a friction force, causing precession to change the alignment.
The other very important thing to consider is the so-called surface effect. We do not have an infinite-in-volume ensemble, there is something like a bar, see figure \ref{friction}.
The effect of the surface and friction is known to lead to a greater alignment and less precession. Alignment would be in the plane of the bar.
Therefore putting the bar in the plane of $(\hat{B},\hat{y})$, leads to the assumptions made in \cite{Moti:2024a}, and we have only $\theta$ and not $\phi$ to integrate over.

Consider a paramagnetic matter within a constant magnetic field (which we assumed to be tangent to the celestial sphere). We are dealing with classical paramagnetic matter in an external magnetic field, meaning that there are some randomly orientated classical spinning tops interacting classically with magnetic field. As a result of the presence of the magnetic field, both before and after the exposure to the the gravitational wave, they have precession about the magnetic field. The effect of the passage of the gravitational wave is to redistribute tangential components of the spin of gyroscopes as $S_t \rightarrow S_t +\Delta S_t$, where $\Delta S_t$ is given by equation \eqref{Delta-S} and depends on $\Delta\Phi$. The magnetic field is in the tangential surface and the Hamiltonian is proportional to  $\vec{B} \vdot \vec{S}$, so this redistribution changes the partition function and entropy with footprints from $\Delta\Phi$.

\section{Details of Calculations of Subsection \ref{SecIII:Kinetic}} \label{App. B}
\setcounter{equation}{0}

The generalized Boltzmann equation is
\begin{equation}
\pdv{f}{x^{\mu}} \dv{x^{\mu}}{\tau} + \pdv{f}{P^{\mu}} \dv{P^{\mu}}{\tau} + \pdv{f}{S^{\mu\nu}} \dv{S^{\mu\nu}}{\tau} = Q \ .
\label{App-Ex-Boltzman}
\end{equation}
Using the covariant expressions
\begin{align}
& \Ddv{P^\alpha}{\tau} = \dv{P^{\alpha}}{\tau} + \Gamma^{\alpha}_{\mu\nu} U^{\mu} P^{\nu}\\
& \Ddv{S^{\alpha\beta}}{\tau} = \dv{S^{\alpha\beta}}{\tau} + \Gamma^{\alpha}_{\mu\nu} U^{\mu} S^{\nu\beta} + \Gamma^{\beta}_{\mu\nu} U^{\mu} S^{\alpha\nu}
\end{align}
in the equation \eqref{App-Ex-Boltzman}, the Boltzmann equation becomes
\begin{equation}
\pdv{f}{x^{\mu}} U^{\mu} +
\pdv{f}{P^{\mu}} \left( \Ddv{P^{\mu}}{\tau} - \Gamma^{\mu}_{\alpha\beta} U^{\alpha} P^{\beta} \right) + 
\pdv{f}{S^{\mu\nu}} \left(  \Ddv{S^{\mu\nu}}{\tau} - \Gamma^{\mu}_{\alpha\beta} U^{\alpha} S^{\beta\nu} - \Gamma^{\nu}_{\alpha\beta} U^{\alpha} S^{\mu\beta} \right) = Q \ .
\end{equation}

Since the four-velocity of the extended object satisfies
\begin{equation}
U^{\mu} = \dfrac{\bar{Z}}{m^2} \left( P^{\mu} + \tilde{\Pi}^{\mu}   \right);\ \ \ \ \ \ \ \ \tilde{\Pi}^{\mu} = \dfrac{1}{2} ~ ( \dfrac{S^{\mu\nu} R_{\nu\sigma\alpha\beta}P^{\sigma}S^{\alpha\beta}}{m^2 + \dfrac{1}{4}R_{\alpha\beta\gamma\delta}S^{\alpha\beta}S^{\gamma\delta}} )
\end{equation}
one arrives at
\begin{multline}
 \dfrac{\bar{Z}}{m^2} \pdv{f}{x^{\mu}} \left( P^{\mu} + \tilde{\Pi}^{\mu}   \right) +
\pdv{f}{P^{\mu}} \left( \Ddv{P^{\mu}}{\tau} - \dfrac{\bar{Z}}{m^2}  \Gamma^{\mu}_{\alpha\beta} \left( P^{\alpha} + \tilde{\Pi}^{\alpha}   \right) P^{\beta} \right) + \\
\pdv{f}{S^{\mu\nu}} \left(  \Ddv{S^{\mu\nu}}{\tau} - \dfrac{\bar{Z}}{m^2}  \left( \Gamma^{\mu}_{\alpha\beta}  S^{\beta\nu} + \Gamma^{\nu}_{\alpha\beta}  S^{\mu\beta}\right) \left( P^{\alpha} + \tilde{\Pi}^{\alpha}   \right) \right) = Q  \  .
\label{App-3rd-Ed}
\end{multline}
After rearranging and simplifying, the expression takes the final form 
\begin{multline}
 \dfrac{\bar{Z}}{m^2} \left( \pdv{f}{x^{\mu}} P^{\mu}   -
\pdv{f}{P^{\mu}} \Gamma^{\mu}_{\alpha\beta}  P^{\alpha} P^{\beta}  - \pdv{f}{S^{\mu\nu}} \left( \Gamma^{\mu}_{\alpha\beta}  S^{\nu\beta} + \Gamma^{\nu}_{\alpha\beta}  S^{\mu\beta}\right)  P^{\alpha} \right)  \\
+ \dfrac{\bar{Z}}{m^2} \pdv{f}{x^{\mu}} \tilde{\Pi}^{\mu} +  \pdv{f}{P^{\mu}} \Ddv{P^{\mu}}{\tau} + \pdv{f}{S^{\mu\nu}}  \Ddv{S^{\mu\nu}}{\tau}
-  \dfrac{\bar{Z}}{m^2}  \tilde{\Pi}^{\alpha}  \left(  \pdv{f}{P^{\mu}} P^{\beta}  \Gamma^{\mu}_{\alpha\beta}+ \pdv{f}{S^{\mu\nu}} \left( \Gamma^{\mu}_{\alpha\beta}  S^{\nu\beta} + \Gamma^{\nu}_{\alpha\beta}  S^{\mu\beta}\right)   \right)
 = Q   \  .
\end{multline}

The last term on the $l.h.s$ is of the second order in the spacetime perturbations, and thus it can be neglected,
\begin{multline}
\pdv{f}{x^{\mu}} P^{\mu}   -
\pdv{f}{P^{\mu}} \Gamma^{\mu}_{\alpha\beta}  P^{\alpha} P^{\beta}  - \pdv{f}{S^{\mu\nu}} \left( \Gamma^{\mu}_{\alpha\beta}  S^{\nu\beta} + \Gamma^{\nu}_{\alpha\beta}  S^{\mu\beta}\right)  P^{\alpha} = \\
-\pdv{f}{x^{\mu}} \tilde{\Pi}^{\mu} -  \dfrac{m^2}{\bar{Z}} \pdv{f}{P^{\mu}} \Ddv{P^{\mu}}{\tau} - \dfrac{m^2}{\bar{Z}}  \pdv{f}{S^{\mu\nu}}  \Ddv{S^{\mu\nu}}{\tau}  + Q \ .
\label{App-4th-Ed}
\end{multline}

To derive the balance equations, one has to multiply the Boltzmann equation by an appropriate function ($\psi$) and integrate the resulting equation over the entire phase space, i.e. 
\begin{multline}
\int \psi \left[ \pdv{f}{x^{\mu}} P^{\mu}   -
\pdv{f}{P^{i}} \Gamma^{i}_{\alpha\beta}  P^{\alpha} P^{\beta}  - \pdv{f}{S^{\mu\nu}} \left( \Gamma^{\mu}_{\alpha\beta}  S^{\nu\beta} + \Gamma^{\nu}_{\alpha\beta}  S^{\mu\beta}\right)  P^{\alpha} \right] \dfrac{g}{P_0} \dd{\chi}_s \\
= - \int \psi \left[ \pdv{f}{x^{\mu}} \tilde{\Pi}^{\mu} +  \dfrac{m^2}{\bar{Z}} \pdv{f}{P^{i}} \Ddv{P^{i}}{\tau} + \dfrac{m^2}{\bar{Z}}  \pdv{f}{S^{\mu\nu}}  \Ddv{S^{\mu\nu}}{\tau} \right]   \dfrac{g}{P_0}  \dd{\chi}_s + \int \psi \mathcal{Q} \dfrac{g}{P_0} \dd{\chi}_s  \  .
\label{App-MainEq}
\end{multline}

Let's simplify each side separately. For the $l.h.s$
\begin{align}
l.h.s & =  \int \psi  \pdv{f}{x^{\mu}}  P^{\mu} \dfrac{g}{P_0} \dd{\chi}_s -  \int \psi  \pdv{f}{P^{i}} \Gamma^{i}_{\alpha\beta} P^{\alpha} P^{\beta}  \dfrac{g}{P_0} \dd{\chi}_s  -  \int \psi \pdv{f}{S^{\mu\nu}} \left( \Gamma^{\mu}_{\alpha\beta}  S^{\nu\beta} + \Gamma^{\nu}_{\alpha\beta}  S^{\mu\beta}\right)  P^{\alpha} \dfrac{g}{P_0} \dd{\chi}_s  \nonumber \\
& =  \int \pdv{ }{x^{\mu}}\left(f \psi  P^{\mu}   \dfrac{g}{P_0} \right)  \dd{\chi}_s -  \int f	\pdv{\psi}{x^{\mu}}  P^{\mu}  \dfrac{g}{P_0} \dd{\chi}_s -  \underbrace{\int f \psi \pdv{ }{x^{\mu}} \left( P^{\mu}\dfrac{g}{P_0} \right)  \dd{\chi}_s }_{\text{LT1}}  \nonumber \\
&  -  \underbrace{\int \pdv{ }{P^{i}}\left( f \psi  \Gamma^{i}_{\mu\nu} P^{\mu} P^{\nu}  \dfrac{g}{P_0} \right) \dd{\chi}_s }_{0}  + \int f \pdv{\psi}{P^{i}}  \Gamma^{i}_{\mu\nu} P^{\mu} P^{\nu}  \dfrac{g}{P_0} \dd{\chi}_s   + \underbrace{\int  f \psi \pdv{ }{P^{i}} \left(  \Gamma^{i}_{\mu\nu} P^{\mu} P^{\nu}  \dfrac{g}{P_0} \right) \dd{\chi}_s }_{\text{LT2}} \nonumber \\
 &  -  \underbrace{\int \pdv{}{S^{\mu\nu}} \left(  f \psi \left( \Gamma^{\mu}_{\alpha\beta}  S^{\beta\nu} + \Gamma^{\nu}_{\alpha\beta}  S^{\mu\beta}\right)  P^{\alpha} \dfrac{g}{P_0} \right) \dd{\chi}_s }_{0} + \int f \pdv{\psi}{S^{\mu\nu}}\left( \Gamma^{\mu}_{\alpha\beta}  S^{\nu\beta} + \Gamma^{\nu}_{\alpha\beta}  S^{\mu\beta}\right)  P^{\alpha} \dfrac{g}{P_0}  \dd{\chi}_s  \nonumber \\
&   + \underbrace{ \int f \psi   \pdv{ }{S^{\mu\nu}} \left(  \left( \Gamma^{\mu}_{\alpha\beta}  S^{\nu\beta} + \Gamma^{\nu}_{\alpha\beta}  S^{\mu\beta}\right)  P^{\alpha} \dfrac{g}{P_0} \right) \dd{\chi}_s }_{\text{LT3}}
\end{align}
where LT1+LT2+LT3 vanishes because of the Liouville Theorem. This helps us to have a more compact form
\begin{align}
l.h.s  & = \int \pdv{ }{x^{\mu}}\left(f \psi  P^{\mu} \dfrac{g}{P_0} \right)  \dd{\chi}_s \nonumber \\
& -  \int f \pdv{\psi}{x^{\mu}}  P^{\mu}  \dfrac{g}{P_0} \dd{\chi}_s  + \int f  \pdv{\psi}{P^{i}} \Gamma^{i}_{\mu\nu} P^{\mu} P^{\nu} \dfrac{g}{P_0} \dd{\chi}_s  +  \int  f \pdv{\psi}{S^{\mu\nu}}  \left( \Gamma^{\mu}_{\alpha\beta}  S^{\nu\beta} + \Gamma^{\nu}_{\alpha\beta}  S^{\mu\beta}\right)  P^{\alpha} \dfrac{g}{P_0}  \dd{\chi}_s \nonumber \\
& = \int \pdv{}{x^{\mu}}\left( f  \psi P^{\mu}   \sqrt{-g} \dfrac{\dd[3]{P}}{P_0} \dd{\Omega}_s \right) \sqrt{-g}  \dd[4]{x} + \Gamma^{\kappa}_{\mu\kappa} \int  \left( f \psi P^{\mu}   \sqrt{-g} \dfrac{\dd[3]{P}}{P_0} \dd{\Omega}_s \right) \sqrt{-g}  \dd[4]{x} \nonumber \\
& -  \int  f \pdv{\psi}{x^{\mu}}  P^{\mu} \dfrac{g}{P_0} \dd{\chi}_s  + \int  f \pdv{\psi}{P^{i}} \Gamma^{i}_{\mu\nu} P^{\mu} P^{\nu} \dfrac{g}{P_0} \dd{\chi}_s  +  \int f \pdv{\psi}{S^{\mu\nu}}  \left( \Gamma^{\mu}_{\alpha\beta}  S^{\nu\beta} + \Gamma^{\nu}_{\alpha\beta}  S^{\mu\beta}\right)  P^{\alpha} \dfrac{g}{P_0}  \dd{\chi}_s 
\end{align}
which simplifies to
\begin{align}
l.h.s & = \int \sqrt{-g} \dd[4]{x} \left[ \nabla_{\mu} \int \left( f  \psi P^{\mu} \sqrt{-g} \dfrac{\dd[3]{P} }{P_0} \right)  \right. \nonumber \\
& \left. -  \int  f \left( \pdv{\psi}{x^{\mu}}  P^{\mu} \sqrt{-g}  - \pdv{\psi}{P^{i}}  \Gamma^{i}_{\mu\nu} P^{\mu} P^{\nu} \sqrt{-g} -  \pdv{\psi}{S^{\mu\nu}} \left( \Gamma^{\mu}_{\alpha\beta}  S^{\nu\beta} + \Gamma^{\nu}_{\alpha\beta}  S^{\mu\beta}\right) P^{\alpha} \sqrt{-g} \right) \sqrt{-g} \dfrac{\dd[3]P}{P_0}  \dd{\Omega}_s \right] \ .
\label{App-LHS}
\end{align}
One the other hand, the $r.h.s$ of \eqref{App-MainEq} is
\begin{equation}
r.h.s = - \int \psi \left[ \pdv{f}{x^{\mu}} \tilde{\Pi}^{\mu} +  \dfrac{m^2}{\bar{Z}} \pdv{f}{P^{i}} \Ddv{P^{i}}{\tau} + \dfrac{m^2}{\bar{Z}}  \pdv{f}{S^{\mu\nu}}  \Ddv{S^{\mu\nu}}{\tau} \right]   \dfrac{g}{P_0}  \dd{\chi}_s + \int \psi \mathcal{Q} \dfrac{g}{P_0} \dd{\chi}_s  \  .
\end{equation}
The MPD equations \eqref{MP-Momentum}-\eqref{MP-Velocity} restrict the spin evolution to 
\begin{equation}
\Ddv{S^{\mu\nu}}{\tau}  = -\dfrac{1}{m^2} \left( S^{\mu\kappa} P^{\nu} + S^{\kappa\nu} P^{\mu}\right) \Ddv{P_{\kappa}}{\tau} \ .
\end{equation}
Then
\begin{equation}
r.h.s = - \int \psi \left[ \pdv{f}{x^{\mu}} \tilde{\Pi}^{\mu} +  \dfrac{m^2}{\bar{Z}} \left( \pdv{f}{P^{\mu}} g^{\mu\kappa} - \dfrac{1}{m^2}  \pdv{f}{S^{\mu\nu}}  \left(  S^{\mu\kappa} P^{\nu} + S^{\kappa\nu} P^{\mu} \right)    \right) \Ddv{P_{\kappa}}{\tau}   \right]   \dfrac{g}{P_0}  \dd{\chi}_s + \int \psi \mathcal{Q} \dfrac{g}{P_0} \dd{\chi}_s
\end{equation}
and thus the equation \eqref{App-MainEq} can be rewritten as
\begin{align}
& \int \sqrt{-g} \dd[4]{x} \left[ \nabla_{\mu} \int \left( f \psi P^{\mu} \sqrt{-g} \dfrac{\dd[3]{P} }{P_0}  \dd{\Omega}_s \right)  \right. \nonumber \\
& \left. -  \int f \left(  \pdv{\psi}{x^{\mu}}  P^{\mu}    -   \pdv{\psi}{P^{i}}  \Gamma^{i}_{\mu\nu} P^{\mu} P^{\nu}  -  \pdv{\psi}{S^{\mu\nu}} \left( \Gamma^{\mu}_{\alpha\beta}  S^{\nu\beta} + \Gamma^{\nu}_{\alpha\beta}  S^{\mu\beta}\right) P^{\alpha} \right) \sqrt{-g} \dfrac{\dd[3]P}{P_0}  \dd{\Omega}_s \right] \nonumber \\
& =  \int \sqrt{-g} \dd[4]{x} \left( \int \mathcal{D}_{\text{MPD}} \sqrt{-g} \dfrac{\dd[3]{P}}{P_0} \dd{\Omega}_s  + \int \psi \mathcal{Q} \sqrt{-g} \dfrac{\dd[3]{P}}{P_0} \dd{\Omega}_s  \right)
\label{App-MainEq2}
\end{align}
in which $ \mathcal{D}_{\text{MPD}}$ is defined in equation \eqref{D-MPD} and clearly carries the effects of MPD dynamics.

Now, as stated in the text, we need to compare the term $\nabla_{\mu} \left( \int  \psi f P^{\mu} \sqrt{-g} \dfrac{\dd[3]{P} }{P_0}  \dd{\Omega}_s \right)$  at two points $X$ and $X+\Delta$. Thus, one should calculate
\begin{align}
& \left[ \int \left( \bar{f} \bar{\psi}  P^{\mu} \sqrt{-g} \dfrac{\dd[3]{P} }{P_0}  \dd{\Omega}_s \right)  \right]_{;\mu} -\left[ \int \left( f \psi  P^{\mu} \sqrt{-g} \dfrac{\dd[3]{P} }{P_0}  \dd{\Omega}_s \right)  \right]_{;\mu} \nonumber \\
& =  \int  \left(  \bar{f} \pdv{\bar{\psi}}{x^{\mu}} -  f \pdv{\psi}{x^{\mu}} \right) P^{\mu}\sqrt{-g} \dfrac{\dd[3]P}{P_0}  \dd{\Omega}_s  - 
\int \left( \bar{f} \pdv{\bar{\psi}}{P^{i}}   \bar{\Gamma}^{i}_{\mu\nu} -  f \pdv{\psi}{P^{i}}   \Gamma^{i}_{\mu\nu}  \right) \sqrt{-g} P^{\mu} P^{\nu}  \dfrac{\dd[3]P}{P_0}  \dd{\Omega}_s \nonumber \\
& - \int \left[ \bar{f} \pdv{\bar{\psi}}{S^{\mu\nu}} \left( \bar{\Gamma}^{\mu}_{\alpha\beta}  S^{\nu\beta} + \bar{\Gamma}^{\nu}_{\alpha\beta}  S^{\mu\beta} \right) - f \pdv{\psi}{S^{\mu\nu}} \left( \Gamma^{\mu}_{\alpha\beta}  S^{\nu\beta} + \Gamma^{\nu}_{\alpha\beta}  S^{\mu\beta} \right) \right]  P^{\alpha}  \sqrt{-g} \dfrac{\dd[3]P}{P_0}  \dd{\Omega}_s  \nonumber \\
& + \int \left(\bar{\mathcal{D}}_{\text{MPD}}-\mathcal{D}_{\text{MPD}} \right) \sqrt{-g} \dfrac{\dd[3]{P}}{P_0} \dd{\Omega}_s + \int \left(\bar{\psi} \bar{\mathcal{Q}}-\psi \mathcal{Q} \right) \sqrt{-g} \dfrac{\dd[3]{P}}{P_0} \dd{\Omega}_s \ .
\label{App-De-Main}
\end{align}
Attention to the terms in the last line is needed. The first term, which directly encodes the MPD dynamics, vanishes at leading order due to the linear perturbations of $\psi$ and $f$. The second term also vanishes, since we considered the same collision-less assumption as for the case of point particles.

Then, due to the definitions of $\mathcal{N}^{\mu}_{\BM_{s}}$, $\mathcal{T}^{\mu}_{\BM_{s}}$, $\zeta^{\mu}_{\BM_{s}}$, $\Sigma^{\mu}_{\BM_{s}}$, and $\mathcal{L}^{\mu}_{\BM_{s}}$ in equations \eqref{N-MPD-Mom}-\eqref{Ent-MPD-Mom}, \eqref{S-MPD-Mom} and \eqref{L-MPD-Mom} respectively, the related balance equations are as follows:
\begin{itemize}
\item {\bf Number current conservation}: Substituting $\psi = c$ and $\bar{\psi} = c$ into \eqref{App-De-Main} the $l.h.s$ is $\nabla_{\mu}\bar{\mathcal{N}}^{\mu}_{\BM_s} -\nabla_{\mu}\mathcal{N}^{\mu}_{\BM_s}$ while the $r.h.s$ vanishes because we choose $\psi$ and $\bar{\psi}$ as a constant $c$. Then,
\begin{equation}
\Delta(\nabla_{\mu}\mathcal{N}^{\mu}_{\BM_s}) \equiv \nabla_{\mu}\bar{\mathcal{N}}^{\mu}_{\BM_s} -\nabla_{\mu}\mathcal{N}^{\mu}_{\BM_s}  = \mathcal{O} (2) \ .
\end{equation}

\item {\bf Energy--momentum conservation}: For four-vectors  $\psi^{\nu} = c P^{\nu}$ and $\bar{\psi}^{\nu} = c P^{\nu}$, since the covariant extended form of \eqref{MainEq2} is
\begin{multline}
 \nabla_{\mu} \int \left( \psi^{\nu} f P^{\mu} \sqrt{-g} \dfrac{\dd[3]{P} }{P_0}  \dd{\Omega}_s \right)  = \\
  \int \left( f \left(\nabla_{\mu}\psi^{\nu} \right) P^{\mu}    - f \pdv{\psi^{\nu}}{P^{i}}  \Gamma^{i}_{\alpha\beta} P^{\alpha} P^{\beta}  -  f \pdv{\psi^{\nu}}{S^{\rho\sigma}}   \left( \Gamma^{\rho}_{\alpha\beta}  S^{\sigma\beta} + \Gamma^{\sigma}_{\alpha\beta}  S^{\rho\beta}\right) P^{\alpha} \right) \sqrt{-g} \dfrac{\dd[3]P}{P_0}  \dd{\Omega}_s  \\
+ \int \mathcal{D}_{\text{MPD}}^{\nu} \sqrt{-g} \dfrac{\dd[3]{P}}{P_0} \dd{\Omega}_s  + \int \psi^{\nu} \mathcal{Q} \sqrt{-g} \dfrac{\dd[3]{P}}{P_0} \dd{\Omega}_s ,
\label{App-Cov-Bolt-Ave}
\end{multline}
with
\begin{equation}
\mathcal{D}_{\text{MPD}}^{\nu} \equiv -\psi^{\nu} \left[ \pdv{f}{x^{\mu}} \tilde{\Pi}^{\mu} +  \dfrac{m^2}{\bar{Z}} \left( \pdv{f}{P^{i}} g^{i\kappa} - \dfrac{1}{m^2}  \pdv{f}{S^{\rho\sigma}}  \left(  S^{\rho\kappa} P^{\sigma} + S^{\kappa\sigma} P^{\rho} \right)    \right) \Ddv{\tilde{P}_{\kappa}}{\tau}   \right]  \ .
\end{equation}
Comparison of this equation at two different spacetime points, leads to
\begin{align}
& \left[ \int \left( \bar{f} \bar{\psi}^{\nu}  P^{\mu} \sqrt{-g} \dfrac{\dd[3]{P} }{P_0}  \dd{\Omega}_s \right)  \right]_{;\mu} -\left[ \int \left( f \psi^{\nu}  P^{\mu} \sqrt{-g} \dfrac{\dd[3]{P} }{P_0}  \dd{\Omega}_s \right)  \right]_{;\mu} \nonumber \\
& =  \int \left( \bar{f} \nabla_{\mu} \bar{\psi}^{\nu} - f \nabla_{\mu}\psi^{\nu} \right)  P^{\mu} \sqrt{-g} \dfrac{\dd[3]P}{P_0}  \dd{\Omega}_s  - 
\int \left( \bar{f} \pdv{\bar{\psi}^{\nu}}{P^{i}}   \bar{\Gamma}^{i}_{\alpha\beta} - f  \pdv{\psi^{\nu}}{P^{i}}  \Gamma^{i}_{\alpha\beta}  \right) \sqrt{-g} P^{\alpha} P^{\beta}  \dfrac{\dd[3]P}{P_0}  \dd{\Omega}_s \nonumber \\
& - \int \left[ \bar{f} \pdv{\bar{\psi}^{\nu}}{S^{\rho\sigma}} \left( \bar{\Gamma}^{\rho}_{\alpha\beta}  S^{\sigma\beta} + \bar{\Gamma}^{\sigma}_{\alpha\beta}  S^{\rho\beta} \right) - f \pdv{\psi^{\nu}}{S^{\rho\sigma}} \left( \Gamma^{\rho}_{\alpha\beta}  S^{\sigma\beta} + \Gamma^{\sigma}_{\alpha\beta}  S^{\rho\beta} \right)  \right] P^{\alpha}  \sqrt{-g} \dfrac{\dd[3]P}{P_0}  \dd{\Omega}_s  \nonumber \\
& + \int \left(\bar{\mathcal{D}}_{\text{MPD}}^{\nu}-\mathcal{D}_{\text{MPD}}^{\nu} \right) \sqrt{-g} \dfrac{\dd[3]{P}}{P_0} \dd{\Omega}_s  + \int \left(\bar{\psi}^{\nu} \bar{\mathcal{Q}}-\psi^{\nu} \mathcal{Q} \right) \sqrt{-g} \dfrac{\dd[3]{P}}{P_0} \dd{\Omega}_s
\end{align}
Then, using the definitions of  $\mathcal{T}^{\mu\nu}_{\BM_s}$ and $\bar{\mathcal{T}}^{\mu\nu}_{\BM}$ given in \eqref{psi-T}, we obtain
\begin{align}
\Delta(\nabla_{\mu}\mathcal{T}^{\mu\nu}_{\BM_s}) \equiv \nabla_{\mu}\bar{\mathcal{T}}^{\mu\nu}_{\BM_s} -\nabla_{\mu}\mathcal{T}^{\mu\nu}_{\BM_s} & = \mathcal{O} (2) 
\end{align}

\item {\bf Entropy rate equation}: For $\psi = - ck \ln{\left(\dfrac{f h^3}{e g_s}\right)}$, $\zeta^{\mu}_{\BM_s} = - ck  \int  f \ln{\left(\dfrac{f h^3}{e g_s}\right)}    P^{\mu}  \sqrt{g} \dfrac{ \dd[3]{p} }{p_0}$ and their barred counterparts
we get
\begin{align}
\Delta (\nabla_{\mu} \zeta^{\mu}_{\BM_s}) & =  \int  \left(\bar{f} \partial_{\mu} \bar{\psi} - f \partial_{\mu} \psi  \right)P^{\mu} \sqrt{-g} \dfrac{\dd[3]{P}}{P_0} \dd{\Omega}_s \nonumber \\
& - \int \left( \bar{f} \pdv{\bar{\psi}}{P^{i}}  \ \bar{\Gamma}^{\alpha}_{\mu\nu}  - f \pdv{\psi}{P^{i}}  \ \Gamma^{i}_{\mu\nu}  \right) \sqrt{-g} \dfrac{\dd[3]{P}}{P_0} \dd{\Omega}_s \nonumber\\
& - \int \left[\bar{f}  \pdv{\bar{\psi}}{S^{\mu\nu}}\left( \bar{\Gamma}^{\mu}_{\alpha\beta}  S^{\nu\beta} + \bar{\Gamma}^{\nu}_{\alpha\beta}  S^{\mu\beta} \right)  - f \pdv{\psi}{S^{\mu\nu}} \left( \Gamma^{\mu}_{\alpha\beta}  S^{\nu\beta} + \Gamma^{\nu}_{\alpha\beta}  S^{\mu\beta} \right)  \right] P^{\alpha}  \sqrt{-g} \dfrac{\dd[3]P}{P_0}  \dd{\Omega}_s\nonumber \\
& +  \int \left(\bar{\mathcal{D}}_{\text{MPD}}-\mathcal{D}_{\text{MPD}} \right) \sqrt{-g} \dfrac{\dd[3]{P}}{P_0} \dd{\Omega}_s + \int \left(\bar{\psi} \bar{\mathcal{Q}}-\psi \mathcal{Q} \right) \sqrt{-g} \dfrac{\dd[3]{P}}{P_0} \dd{\Omega}_s
\end{align}
The first two terms on the $r.h.s$ are identical to those in the point particle model and yield $\dfrac{-2k}{3} \left( R^{\beta}_{\ \beta\nu\alpha} + R^{\beta}_{\ \nu \beta\alpha} \right) \Delta^{\alpha} \mathcal{N}^{\rho}_{\BM_{s}} $. The third term vanishes due to the definition of the entropy moment, since $\psi$ carries no spin dependence. The last term also does not contribute at leading order within our collision-less model. Thus,
\begin{equation}
\Delta (\nabla_{\mu} \zeta^{\mu}_{\BM_{s}}) = \nabla_{\mu} \bar{\zeta}^{\mu}_{\BM_{s}}-\nabla_{\mu} \zeta^{\mu}_{\BM_{s}} + \dfrac{2k}{3} \left( R^{\beta}_{\ \beta\nu\alpha} + R^{\beta}_{\ \nu \beta\alpha} \right) \Delta^{\alpha} \mathcal{N}^{\rho}_{\BM_{s}}  > 0
\end{equation}

\item {\bf Spin current equation}: The spin current can be derived from
\begin{equation}
      \Sigma^{\mu\nu \tau}_{\BM_s} =  \displaystyle\int f S^{\mu\nu} P^{\tau}  \sqrt{-g} \dfrac{\dd[3]{P}}{P_0} \dd{\Omega}_s 
\end{equation}
with
\begin{align}
& \psi^{\mu\nu} = cS^{\mu\nu}, \quad \quad 
 \Sigma^{\mu\nu\tau}_{\BM_s} = c \int  f S^{\mu\nu} P^{\tau} \sqrt{-g} \dfrac{\dd[3]{P}}{P_0} \dd{\Omega}_s \nonumber \\
& \bar{\psi}^{\mu\nu} = c\bar{S}^{\mu\nu} , \quad \quad  \bar{\Sigma}^{\mu\nu\tau}_{\BM_s} = c \int \bar{f}  S^{\mu\nu} P^{\tau} \sqrt{-g} \dfrac{\dd[3]{P}}{P_0} \dd{\Omega}_s
\end{align} 
Extending the \eqref{App-MainEq2} for a second-rank tensor, we arrive at
\begin{multline}
 \nabla_{\tau} \int \left( f \psi^{\mu\nu} P^{\tau} \sqrt{-g} \dfrac{\dd[3]{P} }{P_0}  \dd{\Omega}_s \right)  = \\
  \int \left( f \left(\nabla_{\tau}\psi^{\mu\nu} \right) P^{\tau}   - f  \pdv{\psi^{\mu\nu}}{P^{i}}  \Gamma^{i}_{\alpha\beta} P^{\alpha} P^{\beta}  - \pdv{\psi^{\mu\nu}}{S^{\rho\sigma}}  f \left( \Gamma^{\rho}_{\alpha\beta}  S^{\sigma\beta} + \Gamma^{\sigma}_{\alpha\beta}  S^{\rho\beta}\right) P^{\alpha} \right) \sqrt{-g} \dfrac{\dd[3]P}{P_0}  \dd{\Omega}_s  \\
+ \int \mathcal{D}_{\text{MPD}}^{\mu\nu} \sqrt{-g} \dfrac{\dd[3]{P}}{P_0} \dd{\Omega}_s  + \int \psi^{\mu\nu} \mathcal{Q} \sqrt{-g} \dfrac{\dd[3]{P}}{P_0} \dd{\Omega}_s ,
\end{multline}
where
\begin{equation}
\mathcal{D}_{\text{MPD}}^{\mu\nu} \equiv -\psi^{\mu\nu} \left[ \pdv{f}{x^{\lambda}} \tilde{\Pi}^{\lambda} +  \dfrac{m^2}{\bar{Z}} \left( \pdv{f}{P^{i}} g^{i\kappa} - \dfrac{1}{m^2}  \pdv{f}{S^{\rho\sigma}}  \left(  S^{\rho\kappa} P^{\sigma} + S^{\kappa\sigma} P^{\rho} \right)    \right) \Ddv{\tilde{P}_{\kappa}}{\tau}   \right]  \ .
\end{equation}
Following the same process as for the energy--momentum tensor, we get
\begin{align}
& \left[ \int \left( \bar{f} \bar{\psi}^{\mu\nu}  P^{\tau} \sqrt{-g} \dfrac{\dd[3]{P} }{P_0}  \dd{\Omega}_s \right)  \right]_{;\tau} -\left[ \int \left( f \psi^{\mu\nu}  P^{\tau} \sqrt{-g} \dfrac{\dd[3]{P} }{P_0}  \dd{\Omega}_s \right)  \right]_{;\tau} \nonumber \\
& =  \int  \left( \bar{f} \nabla_{\tau} \bar{\psi}^{\mu\nu} - f \nabla_{\tau}\psi^{\mu\nu} \right) P^{\tau} \sqrt{-g} \dfrac{\dd[3]P}{P_0}  \dd{\Omega}_s  - 
\int \left( \bar{f} \pdv{\bar{\psi}^{\mu\nu}}{P^{i}}   \bar{\Gamma}^{i}_{\alpha\beta} -   f \pdv{\psi^{\mu\nu}}{P^{\lambda}} \Gamma^{\lambda}_{\alpha\beta}  \right) \sqrt{-g} P^{\alpha} P^{\beta}  \dfrac{\dd[3]P}{P_0}  \dd{\Omega}_s \nonumber \\
& - \int \left[ \bar{f}  \pdv{\bar{\psi}^{\mu\nu}}{S^{\rho\sigma}}\left( \bar{\Gamma}^{\rho}_{\alpha\beta}  S^{\sigma\beta} + \bar{\Gamma}^{\sigma}_{\alpha\beta}  S^{\rho\beta} \right) - f \pdv{\psi^{\mu\nu}}{S^{\rho\sigma}} \left( \Gamma^{\rho}_{\alpha\beta}  S^{\sigma\beta} + \Gamma^{\sigma}_{\alpha\beta}  S^{\rho\beta} \right)  \right] P^{\alpha}  \sqrt{-g} \dfrac{\dd[3]P}{P_0}  \dd{\Omega}_s  \nonumber \\
& + \int \left(\bar{\mathcal{D}}_{\text{MPD}}^{\mu\nu}-\mathcal{D}_{\text{MPD}}^{\mu\nu} \right) \sqrt{-g} \dfrac{\dd[3]{P}}{P_0} \dd{\Omega}_s  + \int \left(\bar{\psi}^{\mu\nu} \bar{\mathcal{Q}}-\psi^{\mu\nu} \mathcal{Q} \right) \sqrt{-g} \dfrac{\dd[3]{P}}{P_0} \dd{\Omega}_s
\label{App-2nd-rank-der}
\end{align}
For the special choice of $\psi^{\mu\nu} = c S^{\mu\nu}$, after expanding the covariant derivative of the first term at the $r.h.s$ we have
\begin{align}
\Delta(\nabla_{\tau}\Sigma^{\mu\nu\tau}_{\BM_s}) & \equiv \nabla_{\tau}\bar{\Sigma}^{\mu\nu\tau}_{\BM_s} -\nabla_{\tau}\Sigma^{\mu\nu\tau}_{\BM_s} \nonumber \\ 
& =  c \int  \bar{f} \left(  \partial_{\tau} \bar{S}^{\mu\nu}  + \bar{\Gamma}^{\mu}_{\tau\alpha} S^{\alpha\nu} + \bar{\Gamma}^{\nu}_{\tau\alpha} S^{\mu\alpha} \right)  P^{\tau} \sqrt{-g} \dfrac{\dd[3]P}{P_0}  \dd{\Omega}_s \nonumber \\
& -c \int  f \left( \partial_{\tau} S^{\mu\nu}  + \Gamma^{\mu}_{\tau\alpha} S^{\alpha\nu} + \Gamma^{\nu}_{\tau\alpha} S^{\mu\alpha} \right)  P^{\tau} \sqrt{-g} \dfrac{\dd[3]P}{P_0}  \dd{\Omega}_s \nonumber \\
& - c \int \left[ \bar{f} \pdv{\bar{S}^{\mu\nu}}{S^{\rho\sigma}}\left( \bar{\Gamma}^{\rho}_{\alpha\beta}  S^{\sigma\beta} + \bar{\Gamma}^{\sigma}_{\alpha\beta}  S^{\rho\beta} \right) - f \pdv{S^{\mu\nu}}{S^{\rho\sigma}} \left( \Gamma^{\rho}_{\alpha\beta}  S^{\sigma\beta} + \Gamma^{\sigma}_{\alpha\beta}  S^{\rho\beta} \right)  \right] P^{\alpha}  \sqrt{-g} \dfrac{\dd[3]P}{P_0}  \dd{\Omega}_s  \nonumber \\
& + \mathcal{O}(2) + \mathcal{O}(\mathcal{D}_{\text{MPD}}^{\mu\nu}) 
\end{align}
Let's expand the third line, and forget the higher order terms, then
\begin{align}
\Delta(\nabla_{\tau}\Sigma^{\mu\nu\tau}_{\BM_s}) &  =    c \int   \bar{f} \left( \bar{\Gamma}^{\mu}_{\tau\alpha} S^{\alpha\nu} + \bar{\Gamma}^{\nu}_{\tau\alpha} S^{\mu\alpha} \right) P^{\tau} \sqrt{-g} \dfrac{\dd[3]P}{P_0}  \dd{\Omega}_s \nonumber \\
& - c \int f \left( \Gamma^{\mu}_{\tau\alpha} S^{\alpha\nu} + \Gamma^{\nu}_{\tau\alpha} S^{\mu\alpha} \right)  P^{\tau}  \sqrt{-g} \dfrac{\dd[3]P}{P_0}  \dd{\Omega}_s \nonumber \\
& - \dfrac{c}{2}  \int  \bar{f} \left( \delta^{\mu}_{\rho} \delta^{\nu}_{\sigma} - \delta^{\mu}_{\sigma} \delta^{\nu}_{\rho} \right)  \left( \bar{\Gamma}^{\rho}_{\alpha\beta}  S^{\sigma\beta} + \bar{\Gamma}^{\sigma}_{\alpha\beta}  S^{\rho\beta} \right) P^{\alpha}  \sqrt{-g} \dfrac{\dd[3]P}{P_0}  \dd{\Omega}_s  \nonumber \\
& + \dfrac{c}{2} \int f  \left( \delta^{\mu}_{\rho} \delta^{\nu}_{\sigma} - \delta^{\mu}_{\sigma} \delta^{\nu}_{\rho} \right) \left( \Gamma^{\rho}_{\alpha\beta}  S^{\sigma\beta} + \Gamma^{\sigma}_{\alpha\beta}  S^{\rho\beta} \right) P^{\alpha}  \sqrt{-g} \dfrac{\dd[3]P}{P_0}  \dd{\Omega}_s \ .
\end{align}
After straightforward calculation, we have at hand
\begin{equation}
\Delta(\nabla_{\tau}\Sigma^{\mu\nu\tau}_{\BM_s})  =   
 c \int  f \left( - \left( \bar{\Gamma}^{\mu}_{\tau\alpha}- \Gamma^{\mu}_{\tau\alpha} \right) S^{\nu\alpha} +  \left( \bar{\Gamma}^{\nu}_{\tau\alpha}- \Gamma^{\nu}_{\tau\alpha} \right) S^{\mu\alpha} \right) P^{\tau} \sqrt{-g} \dfrac{\dd[3]P}{P_0}  \dd{\Omega}_s
\end{equation}
Since $ \bar{\Gamma}^{\mu}_{\tau\alpha}- \Gamma^{\mu}_{\tau\alpha} = \Delta^{\rho} \partial_{\rho} \Gamma^{\mu}_{\tau\alpha}$, and  $\partial_{\rho} \Gamma^{\mu}_{\tau\alpha} = -\dfrac{2}{3} R^{\mu}_{(\tau\alpha)\rho}$ in the Riemann normal coordinate, then
\begin{equation}
\Delta(\nabla_{\tau}\Sigma^{\mu\nu\tau}_{\BM_s})  =   
\dfrac{4}{3} c \int  f \Delta^{\rho} R^{\left[ \mu \right.}_{\ (\tau\alpha)\rho} S^{\left. \nu \right] \alpha} P^{\tau} \sqrt{-g} \dfrac{\dd[3]P}{P_0}  \dd{\Omega}_s
\end{equation}

\item {\bf Orbital angular momentum current equation}: Finally, for the orbital angular momentum current
\begin{equation}
 L^{\mu\nu\tau}_{\BM_s} =  c \displaystyle\int f \left( X^{\mu}P^{\nu} - X^{\nu} P^{\mu}    \right) P^{\tau}  \sqrt{-g} \dfrac{\dd[3]{P}}{P_0} \dd{\Omega}_s 
\end{equation}
one has to use
\begin{align}
& \psi^{\mu\nu} = c  \left( X^{\mu}P^{\nu} - X^{\nu} P^{\mu}    \right), \qquad L^{\mu\nu\tau}_{\BM_s} =  c \int f \left( X^{\mu}P^{\nu} - X^{\nu} P^{\mu}    \right) P^{\tau}  \sqrt{-g} \dfrac{\dd[3]{P}}{P_0} \dd{\Omega}_s \nonumber \\
& \bar{\psi}^{\mu\nu} = c  \left( \bar{X}^{\mu}P^{\nu} - \bar{X}^{\nu} P^{\mu}    \right), \qquad \bar{L}^{\mu\nu\tau}_{\BM_s} =  c \int \bar{f}  \left( \bar{X}^{\mu}P^{\nu} - \bar{X}^{\nu} P^{\mu}  \right) P^{\tau} \sqrt{-g} \dfrac{\dd[3]{P}}{P_0} \dd{\Omega}_s
\end{align}
in  \eqref{App-2nd-rank-der}, and thus
\begin{align}
\Delta(\nabla_{\tau} L^{\mu\nu\tau}_{\BM_s}) & \equiv \nabla_{\tau}\bar{L}^{\mu\nu\tau}_{\BM_s} -\nabla_{\tau} L^{\mu\nu\tau}_{\BM_s} \nonumber \\
& = c \int   \bar{f} \left(  \partial_{\tau} \bar{L}^{\mu\nu}  + \bar{\Gamma}^{\mu}_{\tau\alpha} \bar{L}^{\alpha\nu} + \bar{\Gamma}^{\nu}_{\tau\alpha} \bar{L}^{\alpha\mu} \right)P^{\tau}  \sqrt{-g} \dfrac{\dd[3]P}{P_0}  \dd{\Omega}_s \nonumber \\
& - c \int f \left( \partial_{\tau} L^{\mu\nu}  + \Gamma^{\mu}_{\tau\alpha} L^{\alpha\nu} + \Gamma^{\nu}_{\tau\alpha} L^{\alpha\mu} \right)  P^{\tau}  \sqrt{-g} \dfrac{\dd[3]P}{P_0}  \dd{\Omega}_s \nonumber \\
&  - c \int \left( \bar{f} \pdv{\bar{L}^{\mu\nu}}{P^{i}} \bar{\Gamma}^{i}_{\alpha\beta}  - f \pdv{L^{\mu\nu}}{P^{i}} \Gamma^{i}_{\alpha\beta} \right) P^{\alpha} P^{\beta}\sqrt{-g} \dfrac{\dd[3]P}{P_0}  \dd{\Omega}_s \nonumber \\
& + \mathcal{O} (2) + \mathcal{O}(\mathcal{D}_{\text{MPD}}^{\mu\nu})  
\end{align}
Considering the definitions of $L^{\alpha\beta}$ and $\bar{L}^{\alpha\beta}$ besides that $\bar{X}^{\alpha} = X^{\alpha} + \Delta^{\alpha}$, and $ \bar{\Gamma}^{\tau}_{\mu\nu}  = \Gamma^{\tau}_{\mu\nu} + \Delta^{\rho} \partial_{\rho} \Gamma^{\tau}_{\mu\nu}$, we get
\begin{align}
\Delta(\nabla_{\tau} L^{\mu\nu\tau}_{\BM_s}) & = c \int   \bar{f} \left( \bar{\Gamma}^{\mu}_{\tau\alpha} \left( L^{\alpha\nu} + \Delta^{\alpha} P^{\nu} -\Delta^{\nu} P^{\alpha} \right) + \bar{\Gamma}^{\nu}_{\tau\alpha} \left( L^{\mu\alpha} + \Delta^{\mu} P^{\alpha} -\Delta^{\alpha} P^{\mu} \right)  \right) P^{\tau} \sqrt{-g} \dfrac{\dd[3]P}{P_0}  \dd{\Omega}_s \nonumber \\
& - c \int   f \left( \Gamma^{\mu}_{\tau\alpha} L^{\alpha\nu} + \Gamma^{\nu}_{\tau\alpha} L^{\alpha\mu} \right) P^{\tau} \sqrt{-g} \dfrac{\dd[3]P}{P_0}  \dd{\Omega}_s \nonumber \\
&  - c \int \left(- \bar{f} \left( \bar{X}^{\mu}\delta^{\nu}_{i} - \bar{X}^{\nu}\delta^{\mu}_{i} \right) \bar{\Gamma}^{i}_{\alpha\beta} + f \left( X^{\mu}\delta^{\nu}_{i} - X^{\nu}\delta^{\mu}_{i} \right) \Gamma^{i}_{\alpha\beta}  \right) P^{\alpha} P^{\beta}\sqrt{-g} \dfrac{\dd[3]P}{P_0}  \dd{\Omega}_s
\end{align}
which after simplification, reduces to
\begin{equation}
\Delta(\nabla_{\tau} L^{\mu\nu\tau}_{\BM_s}) = -2c \int f \left( \Delta^{\rho} \partial_{\rho} \Gamma^{\left[ \mu \right.}_{\tau\alpha}  L^{\left. \nu \right] \alpha } -  \Delta^{\alpha}  \Gamma^{\left[\mu\right.}_{\tau\alpha} P^{\left. \nu \right]} \right)  P^{\tau} \sqrt{-g} \dfrac{\dd[3]P}{P_0}  \dd{\Omega}_s
\end{equation}
And at last, in the Riemann normal coordinate, we have
\begin{equation}
\Delta(\nabla_{\tau} L^{\mu\nu\tau}_{\BM_s}) = \dfrac{4}{3} c \int f  \Delta^{\rho}   R^{\left[ \mu \right. }_{\ (\tau\alpha)\rho} L^{\left.  \nu \right] \alpha } P^{\tau} \sqrt{-g} \dfrac{\dd[3]P}{P_0}  \dd{\Omega}_s
\end{equation}
\end{itemize}

\end{document}